\documentstyle[12pt]{article}
\begin{document}
\title{Emergence of Rules in Cell Society: Differentiation, Hierarchy, and Stability}
\author{ Chikara Furusawa and
        Kunihiko Kaneko \\
        {\small \sl Department of Pure and Applied Sciences}\\
        {\small \sl University of Tokyo, Komaba, Meguro-ku, Tokyo 153, JAPAN}
\\}
\date{}
\maketitle
\begin{abstract}

A dynamic model for cell differentiation is studied, where cells with
internal chemical reaction dynamics interact with each other and
replicate.  It leads to spontaneous differentiation of cells
and determination, as is discussed in the isologous diversification.
Following features of the differentiation are obtained:
(1)Hierarchical differentiation from a ``stem'' cell to other cell types, 
with the emergence of the interaction-dependent rules for differentiation;
(2)Global stability of an ensemble of cells consisting of several cell types,
that is sustained by the emergent, autonomous control on the rate of differentiation;
(3)Existence of several cell colonies with different cell-type distributions.
The results provide a novel viewpoint on the origin of complex cell society,
while relevance to some biological problems,
especially to the hemopoietic system, is also discussed.

\end{abstract}

\section{Introduction}

A multicellular organism is an ordered clone of a fertilized egg.
All the cells contain the same genome set but are specialized in different 
ways. The emergence of different cell types is determined rather precisely, 
while the developmental process of the cells, viewed as a cell society, has robustness against perturbations.

In molecular biology, the differentiation processes are often regarded as 
on-off switching processes.  Switch depends on 
inputs by signal molecules, which leads to a variety of cell types
as outputs. 
A large number of reactions between inputs and outputs are represented as 
a ``cascade", where the reactions are assumed to be approximately independent 
of the other reaction processes in the cell.
The switching behavior (given by the sigmoidal function) is assumed to be 
generated from a chain of these reactions.
With this viewpoint, one can decompose the differentiation by
successive local elementary processes.  It enables us
to elucidate the differentiation processes by experimental methods, where 
several signal molecules and essential genes for differentiations are 
identified.
 
Of course, the development progresses through
cooperation of several processes.  Successive
differentiation processes are often expressed as ``canalization", where
differentiations are captured as a result of dynamics
of complex  chemical networks, in contrast with
a linear combination of simple pathways of chemical reactions.
The pioneering study of Kaufmann (1969) demonstrated that
the Boolean network of genes gives a
variety of final states depending on the initial conditions,
and he has suggested that each final state corresponds to each cell 
type. However, needless
to say, a single initial state embedded in a fertilized egg can produce
several different cell types.  Thus, the following questions remain
unanswered about the gene network: How do the different initial states leading to the 
different cell types arise in the process of development?
How does a selection of specific initial  conditions lead to
precise rules of differentiations?  

It should be noted that in the gene network picture
cellular interactions are not explicitly taken into 
account, which should be important in the course of development.
A pioneering study for the pattern formation is due to Turing (1952),
where dynamic instability by the cellular interactions leads to
the pattern formation (Newman, 1990).
However, it remains still unsolved how
such cell-to-cell interactions are incorporated with
internal dynamical complexity including the gene networks (see also Bignone, 1993; Mjolsness et al., 1991; and Thomas et al, 1995).


Hence it is necessary and important to consider a
system of internal dynamics with suitable cell-cell interactions.
One of the authors (KK) and Yomo have performed several simulations of
interacting cells with internal biochemical networks and cell divisions
that lead to the change in the number of degrees of freedom.
The ``isologous diversification theory" is proposed as a general
mechanism of spontaneous differentiation of replicating biological units (Kaneko \& Yomo, 1994, 95, 97).
In the theory, the following three points are essential.

\begin{itemize}

\item
{\bf Spontaneous differentiation:}
The cells, which have oscillatory chemical reactions within, 
differentiate through interaction with other cells.
This differentiation is provided by the separation of orbits in the phase 
space. The dynamics of cells first split into groups with different 
phases of oscillations, and then to groups with different compositions of chemicals.
These differentiations are not caused by specific substances, but are triggered by the instability brought about by nonlinear systems.
The background of this lies in the dynamic clustering
in globally coupled chaotic systems (Kaneko, 1990, 91, 92).

\item
{\bf Inheritance of the differentiated state to the offspring:}
Each differentiated state of a cell is preserved by the cell division
and transmitted to its offspring.
Chemical composition of a cell is recursively kept with respect to divisions.
Thus a kind of ``memory" is formed, through the transfer of initial 
conditions (e.g., of chemicals). By reproduction, the initial condition of a 
cell is chosen so that
the same cell type is produced at the next generation.

\item
{\bf Global stability:}
Multicellular organism often shows a robustness against some perturbation, such as somatic and other mutations.
An extreme example is seen in
a mutation to triploid in newt, where the cell size becomes
three times, but the total cell number is reduced to one third,
and the final body remains not much affected by the mutation (Fankhauser, 1995).

The distribution of cell types obtained is robust against external 
perturbations.
For instance, when the number of one type of cells is decreased by external
removal, the distribution is recovered by further differentiations to generate the removed cell-type.
In this theory, although the instability triggers the differentiations, 
the cell society as a whole is stabilized through
cell-to-cell interactions.

\end{itemize}

In the present paper we extend these previous studies (Kaneko \& Yomo, 1997) to incorporate the formation of a complex cell society.
We extend our model to allow for complex internal dynamics,
in particular, focusing on the following three topics.

\begin{itemize}

\item
{\bf (i) The hierarchical organization and the emergence of stochastic rules;}
The cell differentiation process in nature follows a hierarchical organization.
For example, the pluripotent cells like stem cells give rise to 
committed cells, which further differentiate to 
terminally differentiated cells.  Here the rules of differentiation
are written as expressions of DNA in principle, but it should be noted that the differentiation is often interaction dependent.  Furthermore, in the hemopoietic system, the differentiation process appears to be stochastic, and the probability of each choice
seems to depend also on the distribution of cell types (Ogawa, 1993).
Hence it is interesting how such interaction-dependent rules of
hierarchical differentiation are formed naturally through the
interplay between internal dynamics and cell-to-cell interactions.

\item
{\bf  (ii) Stability of cell types and cell groups;}
Cells belonging to the same cell-type also slightly differ
each other.  Hence discretization of states to types and their continuous change
coexist among cells.
The differentiation rules of cell types are written for 
the discrete types.  When 
cell differentiation is determined, memory of the discrete state
is stable against cell division. Then, the state has to be dynamically
stable (like an
attractor), while for stem cells or undetermined cells,
their state must have both stability and variability (differentiability)
by divisions.  
Here we are interested  in how such stability and differentiability
are compatible as a state of dynamical systems.
Besides the stability of a cellular state,  
stability about the distribution of cell types has to be 
attained through the developmental process. 
For example, the distribution of cell types in the hemopoietic system is robust against external perturbations.
Here we try to answer the question of stability through an interplay of internal dynamics
and cell-cell interaction, that leads to
modulation of internal cellular states and to stability of distribution of the cells
at the ensemble level.

\item
{\bf (iii) Differentiation of cell colony;}
In an organism, there often appears a higher level of differentiations,
leading to several distinct types of tissues.  They consist of
different types of cells and/or different distribution of cell
types.  
Indeed in the hemopoietic system, several colonies consisting of different cell types appear from same stem cells (Nakahata et al., 1982).
It is an important question how a single cell
can form such different cell colonies.  This is a higher level
question than cell differentiation, since the population of
cell types has to be differentiated.

\end{itemize}

In the present paper we study the above three problems
by extending the previous model of cell differentiation,
to allow for complex internal dynamics.  Here the
cellular states are given by a set of chemical concentrations, while
the internal dynamics is given by mutually catalytic reaction
networks.  In contrast with the previous model, the internal dynamics
allows for chaos and also coexistence of multiple attractors.
Interaction among cells is given by the diffusive transport
of chemicals between each cell and a homogeneous environment.
Cell volume is increased through the transport of chemicals 
from the environment, which leads to the cell division
when it is larger than a given threshold.  


By allowing complex dynamics
at the internal cell level, we will show that the above three problems
are answered from our standpoint.
First, hierarchical differentiation of several cell types is formed.
There appears a cell type that plays the role of ``stem cell", from
which different cell types are differentiated.
The probabilistic switch of cell types is given through the internal dynamics,
whose rate is dependent on the interaction, and accordingly, on the
distribution of other cell types. 
Second, the stability of cell types is given as a
``partial attractor" (to be discussed) of the internal dynamics,
stabilized through interactions.  Third, stochastic population dynamics
of cell types emerges as a higher level.  It is found that
this dynamics has several attracting states, which supports 
different stable cell colonies (tissues).

The organization of the paper is as follows.  In \S 2, our model is
presented.  Although a specific type of catalytic reaction network is
adopted in the present paper, it should be noted that the results are
generally seen in a variety of reaction networks.  Although the
interaction we adopt here is global, in the sense that all cells
interact with each other, our main conclusion on the differentiation and
the formation of cell society is invariant even if the ``spatial" effect
is explicitly taken into account as local diffusion process. In \S 3, we
will show numerical results of the evolution of cell society from a
single cell, where the emergence of distinct cell types is given.
Differentiation of these cells is found to obey a specific rule, that
emerges as a higher level than the chemical reaction rules we have
adapted.  The mechanism of discretization of states, and the formation
of (interaction-dependent) cell memory is discussed in \S 4.  The rule
of ``stochastic" differentiation from a stem-type cell is studied in \S
5, where the stability of the population of cell types is noted. In \S
6, the diversity of cell colonies is shown in relation with several
attracting states of a higher-level dynamics, i.e., the population
dynamics of cell types.  Summary and discussions are given in \S7 and \S8, where
relevance of our results to cell biology is discussed, which covers
origin of stem cells, in particular stochastic branching, stability and
diversity of cell colonies in the hemopoietic system, and the origin of
multicellular organism.

\section{ Model}

Our model for differentiation consists of
\begin{itemize}

\item { Internal dynamics by biochemical reaction network within each cell}

\item { Interaction with other cells through media: Inter-cellular  dynamics
}

\item { Cell division}

\end{itemize}

The basic strategy of the modeling follows
the previous works (Kaneko \& Yomo, 1997), although we take different
dynamics for each of the above three processes.
In essence  we assume a network of catalytic reactions for internal dynamics
that allows for a periodic and/or chaotic oscillations of chemicals,  
while the interaction process is just a diffusion of chemicals
through media.

We represent the internal state of a cell by $k$ chemicals' concentrations 
as dynamical variables.  Cells are assumed to be
in surrounding media, where the same set of chemicals is given.
Hence the dynamics of the internal state is represented by  a set of
variables  $x^{(m)} _i(t)$, the concentration of
the {\it m}-th chemical species  at the {\it i}-th cell, at time {\it t}.
The corresponding concentration of the species in the medium is 
represented by a set of variables $X^{(m)} (t)$.
We assume that the medium is well stirred by
neglecting the spatial variation of the concentration, so that all cells interact each other through identical environment.

\subsection{Internal chemical reaction}

Within each cell, there is a network of biochemical reactions.
The network includes not only a complicated metabolic network 
but also reactions associated with genetic expressions, signaling pathways, 
and so on.  In the present model, a cellular state is represented by
the concentrations of $k$ chemicals.

As internal chemical reaction dynamics we choose 
a catalytic network among the $k$ chemicals.
Each reaction from the chemical $i$ to $j$ is assumed to be catalyzed by 
the chemical $\ell$, which
is determined randomly. To represent the reaction-matrix 
we adopt the notation $Con(i,j,\ell)$ which takes unity when the 
reaction from the chemical $i$ to $j$ is catalyzed by $\ell$, and takes 0 otherwise.
Each chemical has several paths to other chemicals, which act as
a substrate to create several enzymes for other reactions.
Thus these reactions form a complicated network.
This matrix is generated randomly before simulations, and is fixed throughout
the simulation.
We use the same reaction-matrix throughout a series of simulations in this paper (see also \S3 and \S7 for dependence on the reaction-matrix).
 
Usually, chemical kinetics with enzymes is solved under some approximations, 
like Michaelis-Menten form.
In this paper, we assume quadratic effect of enzymes.
Thus the reaction from the chemical $m$ to $\ell$ aided by the chemical 
$j$ leads to the term  $e_1 x^{(m)}_i(t) (x^{(j)}_i(t))^2$, where 
$e_1$ is a coefficient for chemical reactions, which is taken identical 
for all paths.
The quadratic effect of enzymes is not essential to our scenario of cell 
differentiations. Several other forms on the internal dynamics 
lead to qualitatively the same behavior, as long as nonlinear oscillation is included.
The scenario of the differentiation which we propose here is independent of 
the details of this specific choice of biochemical dynamics.

Besides the change of chemical concentrations,
we have to take into account the change of the volume of cell.
The volume is now treated as a dynamical variable, which increases as a 
result of transportation of chemicals into the cell from the environment.
Of course, the concentrations of chemicals are diluted according to the 
increase of the volume of the cell.
For simplicity, we assume that the volume of cell is proportional to the sum 
of chemicals in the cell.
Under this assumption, the operation which compensates the concentration of 
chemicals with the volume change is identical to imposing the restriction 
$\sum_{\ell} x^{(\ell)}_i=1$, namely normalizing the chemical concentrations at each step of the 
calculation, while the volume change is calculated from the
transport as will be given later.

\subsection{Interaction with other cells through media}
 
Each cell communicates with its environment through transport of chemicals.
Interactions between cells, thus, occur through the environment.
Here, the environment does not mean external environment for individual organism, but is intended as interstitial environment of each cell.
In this model, we consider only diffusion process through the cell membrane.
Thus, the rates of chemicals transported into a cell are proportional 
to differences of chemical concentrations between the inside and the
outside of the cell.
Of course, the transport through the membrane is not so simple,
including several mechanisms such as channel proteins and endocytosis.
We omit these complicated mechanisms for simplicity.

The transportation or diffusion coefficient should be
different for different chemicals.  Here we assume that
there are two types of chemicals, those which can penetrate the membrane and which can not.
We use the notation $\sigma_m$, which takes 1 if the chemical $x^{(m)}_i$ is penetrable, and 0 otherwise.

To sum up all these process, the dynamics of chemical concentration in each cell is represented as follows:

\begin{equation}
dx^{(\ell)}_i(t)/dt  =  \delta x^{(\ell)}_i(t) -(1/k) \sum_{l=1}^k\delta x^{(\ell)}_i(t) 
\end{equation}
with 
\begin{eqnarray}
\delta x^{\ell}_i(t)  & =  & \sum_{m,j}Con(m,{\ell},j) \;e_1 \;x^{(m)}_i(t) \;(x^{(j)}_i(t))^2 \nonumber \\
& & - \sum_{m',j'} Con({\ell},m',j') \;e_1 \;x^{({\ell})}_i(t) \;(x^{(j')}_i(t))^2  \nonumber \\
& & + \sigma_{\ell} D(X^{(\ell)} (t) -x^{({\ell})}_i (t))
\end{eqnarray}
where the term with $\sum Con(\cdots)$ represents paths coming into $\ell$ and out of $\ell$ respectively.  The term $\delta x_i^{(\ell)}$ gives the
increment of chemical $\ell$, while the second term in eq.(1) 
gives the constraint of $\sum_{\ell}x^{(\ell)}_i(t)=1$ due to the growth of the volume.
The third term in eq.(2) represents the transport between the medium and the cell, where {\it D} denotes a diffusion constant, which we assume to be identical for all chemicals.
Since the penetrable chemicals in the medium can be consumed with the flow to the cells, we need some flow of chemicals (nutrition) into the medium from the outside.
By denoting the external concentration of these chemicals by $\overline{X}$ and its flow rate per volume of the medium by $f$, the dynamics of 
penetrable chemicals in the medium is written as 

\begin{equation}
dX^{(\ell)}(t)/dt  =f\sigma_{\ell}(\overline{X^{(\ell)}} -X^{(\ell)}(t))-(1/V)\sum_{i=1}^N { \sigma_{\ell}D(X^{(\ell)}(t)-x^{(\ell)}_i(t)) }
\end{equation}
where {\it N} denotes the number of the cells in the environment, and $V$ denotes the volume of the medium in the unit of a cell. 

\subsection{Cell division}

Each cell takes penetrable chemicals from the medium as the nutrient, while
the reactions in the cell transform them to unpenetrable chemicals 
which construct the body of the cell such as membrane and DNA.
As a result of chemical flow, the volume of the cell is increased
by the factor $(1+\sum_{\ell} \delta x^{\ell}_i (t))$  per $dt$.
In the present paper, the cell is assumed to
divide into two almost identical cells when
the volume of the cell is doubled.  
\footnote{In other words, the cell divides 
at the time $t$ when
\begin{equation}
\int_{t_b}^t exp(1+\sum_{\ell} \delta x^{\ell}_i (t') ) dt' =2
\end{equation}
is satisfied since the previous division time $t_b$. }
\footnote{
Embryos fall into two general categories: those in which cell
division is accompanied by growth of the cells back to their former volume
(as mammals and birds); those in which cell division results in cells 1/2
the previous volume (as amphibians). 
Although our model here adopts the division process as in mammals and birds,
we have also confirmed that the present differentiation mechanism also
holds for a model with amphibians-like rules, where cell division makes 
cell volume 1/2, and each cell interacts with neighborhood cells 
like gap junctions.
}
 
The concentrations of chemicals in the daughter cells are almost equal to 
the concentrations of the mother cell.
``Almost" here means that the concentrations of chemicals in a daughter cell are slightly different from the mother's. Each cell has $(1+\epsilon)x^{(l)}$ and $(1-\epsilon)x^{(l)}$ respectively with a small ``noise" $ \epsilon $, a random number with a small amplitude, say over $[-10^{-6}, 10^{-6}]$.
Although the existence of imbalance is essential to the differentiation in our model and in nature, the degree of imbalance itself is not essential to our results to be discussed.
The important feature of our model is the amplification of microscopic differences between the cells through the instability of the internal dynamics.


\subsection{Internal dynamics in single cell}

Before studying the dynamics of cell society, we demonstrate a typical 
behavior of our model by taking only one cell and medium.
In our theory, the fundamental assumption is that the internal dynamics of 
chemicals in the cell shows oscillation as in Fig.1.
In real biological systems, oscillations are observed in some chemical substrates such as Ca, NADH, cyclic AMP, and cyclins (Tyson et al., 1996; Hess et al., 1971; Alberts et al., 1994).
Hence it is natural to postulate such oscillatory dynamics to our model.
The importance of oscillatory dynamics in cellular systems has been pointed out by Goodwin (1963).

The nature of internal dynamics by eqs.(1)-(2) depends on the choice
of the reaction network, in particular on the 
number of paths in the reaction matrix.
When the number of reaction paths is small, cellular dynamics falls into a steady state without oscillation, where a small number of chemicals 
is dominant while other chemicals' concentrations vanish.
On the other hand, when the number of reaction paths is large, many chemicals generate each other. 
Then chemical concentrations take constant values (which are often almost equal).
Only for medium number of reaction paths, non-trivial oscillations
of chemicals appear as in Fig.1.  
We use such network
for our simulation.
It is not easy to estimate the number of paths in real biochemical data, although they may suggest the medium number (3-6) of paths as required in our simulation.

Furthermore, the behavior of dynamics depends on the number of penetrating chemicals.
The number of penetrating chemicals is another control parameter for the capacity of the oscillation or differentiation.
When the number of penetrating chemicals is small, e.g., only one, the rate of randomly chosen reaction networks which show oscillatory dynamics is small.
On the other hand, when the number of penetrating chemicals is too large, it is also difficult to obtain the network with oscillatory dynamics.

Another relevant factor to the nature of internal dynamics is the frequency of auto-catalytic paths.
Indeed, the oscillatory dynamics is rather common as the number of auto-catalytic paths is increased (see \S7).

\section{Differentiation Process: Numerical Results}

We have performed several simulations of our model with different chemical 
networks and different parameters.
Since typical behaviors are rather common, we present our results
by taking a specific chemical network with the number of chemicals $k=20$.
\footnote{
In this example, we do not choose reaction paths equivalently among all chemicals, but select two class of reactions randomly.
One class of reactions is paths from penetrable to any other chemicals, and another is paths from any of chemicals to penetrable ones.
The purpose of this selection is to enhance auto-catalytic reaction loop, and to get oscillatory reaction dynamics easily.
Of course, reaction networks chosen equivalently and randomly can also show the same type of behavior to be discussed in this paper.
As for relationship between auto-catalytic reactions and our scenario for differentiation, see also \S7.
}

As an initial condition, we take a single cell, with
randomly chosen chemical concentrations of $x^{(\ell)}_i$ satisfying $\sum_{\ell} x^{(\ell)}_i  =1$.
In Fig.1, we have plotted a time series of concentration of the chemicals 
in a cell, when only a single cell is in the medium.
This attractor of the internal chemical dynamics is a limit cycle,
whose period is longer than the plotted range in Fig.1. 
We call this state ``attractor-0" or ``type-0" in this paper.
This is the only attractor that is detected from 
randomly chosen initial conditions \footnotemark.

\footnotetext{As will be seen later, there is another attractor as
a single cell state.  However, this attractor is not observed when the
initial condition is randomly chosen; in other words the basin
volume for it is very small.}

With the diffusion term, external chemicals flow into the
cell, which leads to the increase of the volume of the cell.
Thus the cell  is divided into two, with
almost identical chemical concentrations. Chemicals of
the two daughter cells oscillate coherently, with the same dynamical behavior
as the mother cell (i.e., attractor-0).
Successive cell divisions occur simultaneously, and the cell number increases 
as  $1-2-4-8 \cdots$, up to some threshold number.
At this stage, internal dynamics of each cell belongs to the same attractor (i.e., attractor-0), but the oscillations are no longer synchronized.
The microscopic differences introduced at each cell division are
amplified to a macroscopic level through the interaction, which destroys
the phase coherence.

When the number of cells exceeds this threshold value, some cells 
start to show a different type of dynamics.  The threshold number depends on 
the parameters of our model. In the present example, 2 cells start to show a 
different dynamical behavior (as plotted in Fig.2(a)),
when the total cell number becomes 16.
In Fig.2(a), the time series of the chemicals in this cell are plotted.
We call the state as ``partial attractor-1" (or ``type-1" cell).  
We do not call
it an attractor, since the state does not exist as an attractor
of internal dynamics of a single cell.  As will be discussed later,
the stability of the state
is sustained only through the interaction. 
In Fig.3(a), orbits of chemical concentrations are plotted in the phase space during the transition form type-0 to type-1. 
It shows that each attractor occupies distinct regimes in the phase space.
These two types of cells are
clearly distinguishable as digitally distinct states.
Hence we interpret this phenomenon as differentiation.

As the cell number further increases,
another type of cell appears, which we call
type-2 here.  It is again differentiated from the type-0 cell 
(see Fig.2(b) and Fig.3(b)).  
The type-0 cells have potentiality to differentiate to either 
``1" or ``2", while some of the type-``0" cells remain to be of the same 
type by the division.

For some simulations (i.e., for some initial conditions),
the differentiation process stops at this stage, and only three types of
cells coexist.  In many other simulations, however, the
differentiation process continues.
At this stage, hierarchical differentiation occurs.
The cell type ``1" further differentiates into either of three groups
represented as ``3", ``4", or ``5".
The time series of these three types are shown in Fig.2(c)-(e).
The internal dynamics of each type is plotted in a projected phase space
in Fig.4 .
The orbit of type-1 cell itinerates over the three regions corresponding 
to ``3", ``4", and ``5".
For example, Fig.3(c) shows a switch from type-1 to type-3 in the phase space 
by taking a projection different from that in Fig.3(a)(b) (note the difference of scales).
It is also noted that the difference by cell types  is
more clearly distinguishable by chemicals with lower 
concentrations.
 
In the normal course of cell differentiation process (without external operation), cells of the types ``2" and ``1"
reproduce themselves or further differentiate to the other cell types, but 
the offspring never go back to the type-0 cell. Besides the  cell type-2,
the cell types-3, 4, and 5 reproduce themselves
without any further differentiation.  Among these three types,
only the cell type-5 is an attractor by itself, while others
replicate only under the presence of different types of cells.
Indeed, the type-5 is rather special, whose appearance
destabilizes the cell society consisting of ``0", ``1", and ``2".
Once the type-5 cell appears, all the cells will finally be transformed to this
type.
Whether the type-5 cell appears or not depends on the initial condition,
while the cell society without the type keeps diversity of
cell types (see \S 6).

At this stage the differentiation is determined, and cellular memory is
formed as is first discussed in (Kaneko and Yomo, 1997).  
Accordingly we can draw the cell lineage diagram as
shown in Fig.5, where the division 
process with time is represented by the connected line between mother and 
daughter cells while the color in the figure shows the cell type.

The switch of types by differentiations turns out to obey a specific rule.
In Fig.6, we write down an automaton-like representation of
the rule of differentiation.
The node ``0" has three paths; one to itself, and
the others to the nodes ``1" and ``2''.
The path to itself means replication of the same cell type 
through division, while the other paths give the differentiation to the
corresponding cell types.
Fig.6 represents the potentiality of these differentiations.

Note that this differentiation is not induced directly by 
the tiny differences introduced at the division.
The switch from one cell-type to another does not occur simultaneously with 
the division, but occurs later through the interaction among the cells.
This phenomenon is caused by dynamical instability in the total
system consisting of all cells and medium. 
The tiny difference between two daughter cells is amplified to yield 
macroscopic difference through the interaction.  
Our results show that these transitions are not accompanied by the
cell division but occur through cell-to-cell interactions.
This conclusion is consistent with experimental data, where the onset of new 
gene expression is not always accompanied by the cell division.
According to our theory and simulations, the time lag between the
cell division and the onset of new gene expressions depends on
the cell-to-cell interaction, i.e., the surrounding cells.
On the other hand,
change in the number of degrees of freedom by  division 
amplifies the instability in the dynamics of the total system.
When the instability exceeds some threshold, the differentiations start. 
Then, the emergence of another cell type stabilizes the dynamics of
each cell again.  The cell differentiation process in our model is due to
the amplification of tiny differences by orbital instability (transient chaos), while the coexistence of different cell types stabilizes the system.

\section{State Discretization, Hierarchical Organization and Dual Memory}

One might wonder that our definition of types is rather ambiguous and
is not clearly defined. Indeed one can clearly distinguish them by plotting
and comparing the time series and check how these
orbits are separated.
To confirm that  the state in each type is clearly separated,
we introduce the distance between cells in the $k$-dimensional phase space.

Since a cell's state is determined by chemical concentrations in the
present model, the cellular state is represented by an orbit in
the $k$-dimensional phase space.  Here we first consider 
the average position of an orbit for simplicity;
\begin{equation}
\overline{x^{(\ell)}_i}=(1/T)\int x^{(\ell)}_i(t) dt
\end{equation}
As the difference between two cells we adopt the Euclid distance  
\begin{equation}
D_{i,j} \equiv \sqrt{\sum_{\ell}(\overline{x^{(\ell)}_i}-\overline{x^{(\ell)}_j})^2}
\end{equation}
The distance between two cell types is plotted in Table I.
Note that there remains some difference in the same type of cell as mentioned.
However, this difference is clearly much smaller than that 
between different cell types.  
This demonstrates that the differentiated cell types (from ``0" to ``5") are well-defined as ``digitally" distinct states.
Then one might suspect that these different states may be
just a different attractor in each dynamics.  This is not the case.
Except the type-0 and type-5 cells, the state of differentiated cells
is unstable by itself.
When we start the simulation of a single cell with the state of cell type
``1",``2",``3",or ``4"
with the same media (but without any other cells),
the cell is de-differentiated back to the attractor-0.
The states for types ``1'',``2'',``3'', and ``4'' are
stabilized only through the interaction among other cells. 
For example, the existence of type-0 cells
is necessary to keep the stability of cell types ``1" and ``2".

It is also interesting to compare the bifurcation rule of cell types (in Fig.6) with the distance.  If the history of cell lineage reflects on
the distance of cell features, it is expected that for $j=3,4,5$
$ D_{1,j}<D_{0,j}$ or $D_{1,j}<D_{2,j}$ since  the types 3,4,5 are
derived from the type-1 cell.  This is not necessarily true in Table I.
The reason for this discrepancy is due to the insufficiency in the
representation for the distance measured after taking the average.  
As is seen in Fig.4, 
the orbit of the type-1 cell itinerates over the states close to the
type 3,4, and 5.  Hence it is useful to define the minimal
distance by
\begin{equation}
D^{min}_{i,j} \equiv min_t\left(\sqrt{\sum_{\ell}(x^{(\ell)}_i(t)-x^{(\ell)}_j(t))^2}\right)
\end{equation}
where $min_t$ means the minimum over time.
The distance is given in Table II, where one can clearly
see the hierarchical organization of cell types according to the
bifurcation rule of Fig.6.  The distance between two of cell types ``0",``1", and ``2" is 
smaller than that between ``0" or ``2" and ``3",``4",or ``5".  The distance between 
``1" and ``3",``4", or ``5" is much smaller than others.  

Let us reconsider the form of memory using the distance.
First, the memory of cell types is sustained in the internal dynamics modulated
by the interaction.  The memory corresponds to a 
partial attractor stabilized by the interaction. 
Here, the information on
the distribution of cell types is embedded in each internal dynamics.

For example, each internal dynamics is gradually modified
with the change of distribution of other cells.
In Fig.7 we have studied how the dynamics of the type-2 cell
changes when the rate of type-0 cell is varied.
In the simulation, we choose (a stable) cell society consisting
of types ``0", ``2", and ``3", and successively replace a cell of type-0 by type-3.
To avoid the perturbation due to the change in the number of cells, 
we remove the rule of division in the present simulation, 
to fix the number.
As a result of change of the
distribution of cell types (i.e. the fraction of type-0 cells), 
the dynamics of each cell (e.g., of the type 2) is
modulated.
We have plotted the distance $D_{2,2_0}$ where the $2_0$ denotes the cell when the distribution of cells satisfies $(n_0,n_2,n_3)=(23,50,27)$,
the condition at the left-end point of the axis, 
where $n_k$ represents the number of the type-$k$ cell.
The distance $D_{2,2_0}$ increases (roughly linearly) with the decrease of $n_0$, until the further decrease destabilizes 
the cell society and 
the switching of cells to type-5 starts.
The gradual change of  $D_{2,2_0}$ means that the internal cell state
varies according to the cell distribution.  Hence the global information on
the cell distribution is embedded in the internal cellular state.
We note that this information adopts ``analogue" representation, instead of digital one adopted for the distinct cell type.  Hence our cellular system has both analogue and digital memories.




\section{Interaction-dependent rules and stability of cell society}

In Fig.6, we have shown that the automaton-like rule has emerged
without explicit implementation.  The rule is not solely determined by
its cell type.  When there are multiple choices of differentiation process
(as in $``0" \rightarrow ``0"$, or $``0" \rightarrow ``1"$, and $``0" \rightarrow ``2"$)
the rate of each path is
neither fixed nor random, but depends on the number distribution of cell types in the system, embedded in the internal dynamics.
This implies that a higher level dynamics emerges, which controls the rate of 
cell division and differentiation according to the number of each cell 
type.  In other words, the 
dynamics on the number of each cell type $n_0$,$n_1$,.., and $n_5$
can be represented by $\{n_k\}$ ($k=0,\cdots ,5$).  
(This dynamics should be stochastic, since
we have neglected the information on each cellular state and 
reduced it to only the number of cell types).

This dynamics allows for stability at the level of ensemble of cells.
The variety and the 
population distribution of cell types are robust against external perturbations.  
As an example, let us consider the case with three cell types 
(``0" ,``1" ,``2" in Fig.6). When the type-2 cells are removed
to decrease their population,
events of differentiations from ``0" to ``2" are enhanced, 
and the original cell-type distribution is recovered.

In Fig.8, the rate of differentiation from the type-0 cell to others is 
plotted.  In this simulation, to capture the dynamics of the number of 
each cell type, the total number of cells in the medium is fixed (to {\it N}=100 in the present case), by removing the division rule.
As the initial condition, $N$ cells are placed in the medium, where the 
concentration of chemicals in each cell is selected so that they give type-0, 1, or 2 cell.
The switch of cell types is measured when the system settles down to a stable 
distribution of cell types. 
The simulations are repeated by changing the initial distribution of 
cell types 
$(n_0,n_1,n_2)$, to plot the number of the switches from $0$ to others, 
while the final number
of cells for each type is also plotted.  
As in Fig8, the frequency of
switches from the cell type-0 increases almost linearly with $n_0$ when it is larger than approximately 40\%.  With this switch,
the stability of cell distribution around approximately $(n_0, n_1, n_2)=(40, 30, 30)$ is attained.

This kind of robustness at an ensemble level is expected from our isologous 
diversification theory, since the stability of macroscopic characteristics is 
attained in coupled dynamical systems (Kaneko 1992, 94).  
In our case, the macroscopic 
stability is sustained by the 
change of the rate of differentiation from ``0" to other types.
Recall that, the differentiations from ``1" or ``2" to ``0" does not occur (see Fig.6),
even if some of the type ``0" cells are removed \footnotemark.
In the hierarchical structure represented in Fig.6, the cell at an upper 
node behaves as a stem cell, and regulates the distribution of the cells at a lower 
node.  This type of regulation system is often adopted in the real 
multicellular organism (e.g. in the hemopoietic system)(Schofield et al., 1980).
An important point of our result is that the dynamical differentiation 
process always accompanies this kind of regulation process, without any 
sophisticated programs implemented in advance.
This robustness provides a novel viewpoint to understand how the stability of the cell society is maintained in the multicellular organism.

\footnotetext{Transformation from type-2 to type-0 cells occurs 
as a transient process to type-5 cell, which is seen only in the 
case when the type-5 cell appears and starts to dominate the society.}

\section{Differentiation of Colonies}

The automaton rule of Fig.6 does not necessarily mean that all of
these six types of 
cells coexist in a cell society emerged in the course of the development.
Cell groups consisting only of two or three cell 
types can appear: For example, cell groups only of ``0",``1", and ``2" types and of ``0", ``2", and ``4" types are observed.

This implies that the dynamics on the number of cells of each type has also several stable attractors due to the autonomous control of the rate of differentiation.
They correspond to stable distributions of cell types in each cell group.
In other words, there are several possible distributions of cell types when cells are developed from a single cell.
To confirm it, we have performed the following simulations.
First, we initially put one cell whose internal chemical 
concentrations are chosen randomly.
Then the cell society is evolved following the rules of the present 
model, until
the total cell number reaches a given threshold value, when
we stop the simulation and measure the distribution of cell types.
We have repeated this course of simulations for hundred times, starting from different initial conditions.

In Fig.9, the number of initial configurations leading to a cell-type distribution with a given range of $n_2$ is plotted as a histogram, where the number of type-2 cells $n_2$ is measured when the total cell number has reached 300.
Four peaks are clearly visible at $n_2=0$ ,$\sim100$, $\sim$150, and $\sim$220,
which correspond to possible distinct sets of cell distributions.
As mentioned, the possible set of cell types (from ``0" to ``5")
and the temporal ordering of their appearance 
(e.g., 0 $\rightarrow$  1 $\rightarrow $ 2)
are independent of the initial conditions.
However, at the later stage, several types of cell groups emerge depending on the initial conditions.

The most relevant factor to the choice of cell groups
is the ratio of the numbers of differentiated cells 
(i.e. type-1 and type-2 cells) to 
undifferentiated cells (i.e. type-0 cells) at an early stage of development, 
when the first differentiations from ``0'' to ``1'' and ``2'' occur.  
Thus the fate of cell groups is determined at a rather early stage.

Recall that the differentiation rate of cells (each arrow in Fig.6), and accordingly the higher-level dynamics of ${n_k}$ depend on the distribution of cells ${n_k}$.
The result of Fig.9 implies that there are
several attractors on this higher-level dynamics on ${n_k}$.
As discussed in \S 5, an ``attractor" of this higher
level dynamics is stable against perturbations to change the number of cells
of each type.

In Fig.10 we have shown the flow chart of the change of $(n_0,n_1,n_2)$, where
the direction of change of $n_0$ and $n_2$ is represented by the arrow,
starting from the initial distribution given by
$(n_0,n_1,n_2)$  of the corresponding site.  
To draw the figure, 
we adopt the same rules as in Fig.8,
where the total cell number $(n_0+n_1+n_2)$ 
is fixed to 100, and the division rule is removed.
From the 2-dimensional plane, the number of
cells of types 1,3,4, and 5 are given by $100- n_0-n_2$.
The chart shows that cell colonies on the cell-type distribution  
$\{n_0, \cdots n_5\}$ have at least 5 stable states around $(n_0,n_2)$ =(0,0), (38,32), (30,50), (18,58), and (0,78) respectively.
Each state has a basin of attraction, and the corresponding
cell-type distribution is stable against external perturbations,
as is supported by the higher-level dynamics on $\{n_0, \cdots n_5\}$.

The fixed point at $(n_0,n_2)=(0,0)$ (``A" in the figure)
corresponds to a colony consisting only of type-5 cells, while the fixed point denoted by ``B" corresponds to a colony only of 0,1,2, ``C" of 0,2,3, and ``D" of 0,2,4, respectively.
Indeed, these cell-type distributions correspond to the peaks
of Fig.9, respectively.

Still there is a clear difference between the developmental process
from one cell (Fig.9) and the present simulation (Fig.10) with a fixed cell number.  
Some region 
in the plane of Fig.10 cannot be reached by the simulation from a 
single cell.  For example, the state ``E" consisting of 
types 2 and 4 cannot be obtained from the developmental process from
a single cell.  Furthermore, the state ``B", which does not have a large
attraction volume in Fig.10, has the largest probability to be reached
from the developmental process (see Fig.9).  
This discrepancy is caused by the conjunction of
cell number change with the population dynamics of cell types.
Through the change of the number of cells, the population dynamics shifts from
one flow chart of Fig.10 to another with different number of cells.
The organized cell colony from a single cell has such
developmental constraints.

Now the coexistence of several stable
cell colonies is clear.  Depending on the
initial cell condition, different cell colonies are obtained.
The result here means that several 
types of tissues can appear through the interactions among cells. 
This kind of diversity is often observed in a cultivation 
system of a colony of blood cells starting from a stem cell (Nakahata et al., 1982).

\section{Summary}


In the present paper, we have studied a dynamical model to show that a prototype of cell differentiation occurs as a result of
internal dynamics, interaction, and division.
We have made several simulations choosing several chemical
networks, also with a different number of chemical species, and the same scenario for 
cell differentiation is obtained.
Under the same parameters used in the previous example, approximately 5\% of randomly chosen chemical networks show oscillatory behavior, while others fall into fixed points.
Furthermore, approximately 20\% of these oscillatory dynamics are destabilized through the cell division, where some of the cells differentiate following a specific rule like Fig.6.

Some may cast a question why we can select such oscillatory dynamics to draw a general mechanism for differentiation, even if only a few randomly chosen chemical networks are oscillatory.
One reason why only a few reaction networks are oscillatory is that 
we choose reaction paths randomly and with the identical coefficients.
On the other hand, the chemical reaction network of the real biological system 
is more sophisticated through evolutionary process.
For example, there are positive and negative feedback reactions ubiquitously.
This feedback mechanism, in particular auto-catalytic reaction, is important 
to provide oscillatory dynamics which are observed in the real 
biological system (e.g. Ca oscillation).

In the present model with randomly chosen networks, only few reactions have auto-catalytic effects.
By increasing the rate of auto-catalytic reaction paths,
the probability of the network with oscillatory dynamics and differentiation
gets much higher.  For comparison, we have also studied a class of models
where each chemical can catalyze a reaction to generate itself from 
another chemical, besides the ordinary reaction paths determined randomly.
By sampling several reaction networks, we have found that
40\% of the reaction networks has oscillatory dynamics and more than 
20\% of these dynamics are destabilized to show cell differentiation
by cell divisions.

Then, why are such auto-catalytic reactions common?  To make replications
efficiently, some mechanism to amplify reaction by its
product is generally expected at the first stage of life (Eigen and Schuster, 1979). 
Also, auto-catalytic reactions are necessary to add new metabolites
in the metabolic network through the evolutionary process.
Indeed, when novel chemicals are included in the evolutionary process of 
metabolic network, their concentrations must be amplified by the reactions.
This implies that these new chemicals must constitute an auto-catalytic set (see Appendix of Kaneko and Yomo, 1997).

Let us summarize the consequences of our simulations.
First, we have provided a further support for
isologous diversification previously proposed.
Cells are differentiated through the interplay between intra-cellular 
chemical reaction dynamics and the interaction among cells through 
media. As the cell number is increased, the oscillatory dynamics in 
each cell is destabilized and loses synchrony.  Then, 
some of cells change their internal dynamics, which form a group
with a different stable dynamics.
Discrete, differentiated states appear, which are transmitted
to their daughter cells as a memory.
We interpret this phenomenon as (determined) differentiation.

The differentiation in our theory is caused by the
instability in internal dynamics triggered by cell-to-cell interactions.
Microscopically speaking in  biological terms, this may be regarded as a switching process following signal 
molecules from outside of the cell.  Our theory is not inconsistent
with such biological knowledge, but the point in our theory lies in that
such local transition of internal states has also the information on
macroscopic states, i.e., the distribution of cell types.
With this, the  robustness of cell society emerges in spite of
instability in each internal dynamics.

Besides this further support for the isologous diversification,
we have demonstrated the hierarchical cell differentiation,
generation of interaction-dependent rules, and the existence of
distinct cell groups.

{\bf 1) hierarchical organization} 

Differentiation from a stem cell to two different types, and then 
to three types from one of them are observed. Hierarchical rule of differentiation is thus 
generated.  Although the number of cell types and the
rule of differentiation depend on the choice of chemical networks,
generation of a hierarchical rule (written by the tree-type
 diagram as in Fig.6) is generally observed.  

{\bf 2) generation of rules and internal memory reflecting on the environment}
(that is the distribution of other cell types)

These differentiations obey a specific rule, which emerges
from inter- and intra- dynamics.
It is often believed that the rules of the differentiation, which determine 
when, where, and what type of a cell appears in a multicellular 
organism, should be pre-specified as the information on DNA.
We do not deny such role of DNA, but it should be stressed that 
the rules of differentiation and the higher level 
dynamics emerge through interaction of cells with internal dynamics.
As a consequence of our interaction-based approach,
the diversity of cells
and the stability of cell society naturally follow.

The rate of differentiation and reproduction vary with the 
distribution of cell types.
The global stability of the whole system is obtained, which is sustained by 
regulating the rates of the differentiations.

As a coupled dynamical system, the memory of cell types is
given in a state stabilized by interactions.  This state is
not necessarily an attractor as a single cell dynamics, but
is a ``partial attractor" stabilized only in the presence of suitable
interactions provided by the distribution of other cells.
Through the cell divisions and the evolution of the cell society,
the cells choose suitable interactions so that the memory of
their types is preserved.  This is the mechanism of how the
recursivity of cell types is attained, while the global stability of
cell society is assured through the interaction.  Indeed
such partial attractors lose stability and switch to
other cell types when the interaction by the cell distribution is 
not ``suitable''.

It should be noted that 
two types of memory coexist, analogue and digital.
The former gives information on the cell society, i.e., 
the distribution of cell types, while the latter 
gives a distinct internal state on cell differentiation.  We believe that
such dual memory structure is a general feature in a biological system.
In cell biology, the ``analogue" difference
reflecting on the interaction is known as modulation (Alberts et al., 1994).

{\bf 3) formation of higher level dynamics and diversity of cell groups}

The rule of differentiation depends on the number distribution of other
cell types, for which stochastic dynamics at a higher level is formed.
The result provides the first example that a 2-step higher-level dynamics is formed, that is
a colony level from cellular one, that is formed from a
chemical network level.  Here the dynamics of a colony level (i.e., the change of the number of each cell type) is
``stochastic", because the information on the number of cell types
is not complete, where the lower-level information on the
internal state (of chemical concentrations) is discarded.
It is interesting to note that the macroscopic flow chart
on the number of cell types is formed in spite of the stochasticity.

Our result shows that there are several attracting states
for this higher-level dynamics.  
In biological term, this
corresponds to the existence of several cell colonies,
distinguishable by the number distribution of cell types.
These diverse colonies appear from a single stem cell.
Each cell colony is stable, in the sense that 
the original distribution is 
recovered after  perturbations (of not too large size) 
are added on the cell colony, such as elimination of few cells of
one type.

\section{Discussion}

Of course, there has been preceding theories for cell differentiation.
The idea to regard the differentiation as the transition in cellular 
multi-stationarity is traced back to Delbr\"{u}ck (1949), who
proposed a simple bistable reaction network with two metabolic 
chains that are cross-inhibited by their products.  Indeed the epigenetic 
transmission of such stationary states have been reported 
in unicellular organisms (Novick and Wiener, 1957; Sonneborn, 1964).

In general, this multi-stationarity results from positive and negative 
feedbacks in metabolic reaction networks.  This leads to the viewpoint that 
each differentiated cell state is represented by an attractor of 
intra-cellular dynamics, as has been demonstrated by Kauffman (1969) in
his Boolean network. It leads to a variety of stable states (attractors),
depending on the initial conditions.  Here, each cell type corresponds to 
an attractor of internal networks, 
while an external mechanism is required to 
have the transition between these attractors.

Such mechanism is supported by cell-to-cell interaction. Indeed,
a mechanism of interaction-induced differentiation has been proposed
by Turing's pioneering study (Turing, 1952).
Now, a well-known mechanism of external regulation is gradient of morphogen, 
in which the transition depends on the concentration of chemical substances. (See, however, Kaneko and Yomo (submitted), for instability due
to stochastic fluctuation in the threshold mechanism on the gradient
of chemicals).
Another possible mechanism for interaction-induced differentiation
is proposed by Gordon, where the mechanical wave transmits among the cells and 
controls the cell state (Gordon et al., 1994).

Then, combination of the multi-stationary reaction network and the external 
regulation mechanism might be relevant to explain the local differentiations.
However, to understand the complexity and the stability of cell society, an
important question still remains: How is such external regulation mechanism 
regulated? Is another external mechanism required?

Our results provide a distinct, and plausible standpoint for this problem.
Noticing the interplay between intra-cellular dynamics and interaction,
we have proposed a novel concept ``partial attractor'', 
which is stabilized only by cell-to-cell interaction.
Thus, a cell state is not always determined by the attractor of internal 
dynamics, but it also depends on the other cells.
An important consequence of our results is that there is no distinction between 
internal dynamics that determines the cell state and the regulation mechanism 
of differentiation.  Rather, the mechanism for regulation 
is spontaneously accompanied by the multi-stationarity,
because the number of cells with each partial attractor is found to
depend on the circumstance of cell society which sustains them.


A consequence of our theory is `relativity' of determination of a differentiated cell.
Since our cellular state reflects the interaction, 
the rule of differentiation as well as the recursivity may be
affected by the cells around the cell in concern.
A suitable experiment system to distinguish our theory from previous theories is provided
by a hemopoietic stem cell system, where
spatial pattern mechanism a l'a Turing no longer works.
Now we will discuss briefly relevance of our results
to the cell biology.

{\bf Application to biological problems}

Since our model reaction process does not have one-to-one correspondence to 
any existing biochemical network, one cannot make a detailed
prediction on a specific example in biology. However, the
present scenario sheds a new light on some open questions in biology, by providing a coherent viewpoint on them.
Let us discuss two of them.


Since our results provide hierarchical differentiation,
it is interesting to compare them with such an example in cell biology.
A well known example is a hemopoietic system (Ogawa, 1993).
The blood contains many types of cells with different functions, 
while a pluripotent stem cell in the bone marrow gives rise to all classes of 
blood cells.
The hemopoietic system can be viewed as a hierarchy of cells, where pluripotent stem cells differentiate to progenitor cells determined as ancestors of one or a few terminal differentiated blood cell types.
In general, these terminal blood cells have limited lifespans and are produced throughout the life 
of the animal. Thus, to keep a variety of blood cells,
it is important to control the differentiation and proliferation of the stem cell at the higher level of hierarchy.
However, because of the difficulty of identifying the stem cells in the bone 
marrow, the behavior of the pluripotent stem cells in vivo remains especially elusive.
In the experimental result in vitro, even if the cells have been selected to be as homogeneous as possible, there is a remarkable variability in the sizes and often in the characters of the developed colonies (Nakahata et al., 1982). 
Even if two sister cells are taken immediately after a cell division and 
cultured apart under identical conditions, they frequently give rise to 
colonies that contain different types of blood cells or different distribution of the types of cells.

It is often interpreted that the differentiation of
a hemopoietic stem cell is stochastic whose 
probability is controlled by some other control mechanism, by which
the multicellular system as a whole regulates the distribution of
cell types(Till et al., 1964; Ogawa, 1993).
Results of our model provide a novel interpretation of these experiments.
First,  the rules of differentiation are generated through interactions.
Second, the stochastic differentiation of the cells and regulation of
the probability of the differentiations naturally emerge from the cell-to-cell interaction, without imposing any random event or external regulation mechanism.
Third, the diversity of the colonies which have developed from a same type of 
cell is a natural consequence, as a multi-stability of higher-level dynamics.
We note that a single cell with a slightly different initial condition
can lead to a different colony in our simulation.

Through the chemical substrates such as the Interleukins, the complicated interactions among the blood cells are observed experimentally.
It is plausible to assume the dynamical interaction adopted in our theory.

Here, we propose an experiment on the hemopoietic system to make some predictions.
As mentioned, one of the important consequences of our results is that the states of cells are not always determined by the attractor of internal dynamics, but often are sustained by interaction with other cells.
This implies that some types of cells in the hemopoietic system do not correspond to a stable attractor of internal dynamics, but are stabilized by other cells.
The pluripotent stem cell and the terminal differentiated cell, the top and bottom of the hierarchy respectively, seem to correspond to a stable attractor, because they can stand independently of other blood cells.
On the other hand, the progenitor cells can be observed only in the colonies of blood cell.
We expect that the internal dynamics of these progenitor cells is represented by a partial attractor.
Then, these cells can differentiate back to the cell of higher hierarchy when these cells are separated and cultured independently.
To confirm this hypothesis, differences between an isolated blood cell and that surrounded by other cells should be tested experimentally.
We predict that the potentiality of blood cells to differentiate and to proliferate is quite different these two situations, which confirms the importance of the cell-cell interaction in the hemopoietic system.


In general, there is a level of differentiations in cell.
The determined differentiation keeps the memory even if a cell
is transplanted, while some cells can be transdifferentiated (Alberts et al., 1994).
In our dynamical systems representations, such difference can be expressed 
as the distinction between an attractor by the cell itself and the partial 
attractor stabilized by the interaction.    
The merit in our approach is that such levels of differentiation
appear without external implementation, which is
important when one considers the origin of multicellular organisms.
 
There are several types of cells in a multicellular organism. 
In particular, almost all organisms have distinction between the germ cells and the somatic cells.
The appearance of these two types are controlled elaborately by 
complicated interaction among cells in the contemporary multicellular organisms.
However, it is hard to postulate that such a mechanism appeared at the same time
with the emergence of the multicellular organisms.
Our theory provides one possible solution to this problem.
According to our results, differentiation in a group of identical cells 
occurs through the dynamical interaction among cells, as long as the
intra-cellular reaction dynamics can show nonlinear oscillations. 
The differentiation, as well as the stability of such
diverse cells, is a rather natural consequence of
interacting cells.
At the
next stage in the evolution, more complex cell society must have
appeared, where several types of tissues exist as a higher hierarchy,
which interact each other.
Our result about the diversity of cell group shows potentiality that
several types of tissues appear at this stage, based on the dynamical 
interaction among cells.
In our model, we do not take into account of the spatial variation.
Selection of each cell group, thus, depends on the choice of
different initial conditions.
On the other hand, when the spatial information is included in this system, 
it is possible that several types of cell group coexist at a
(spatially) different region.

Of course, contemporary multicellular organisms such as mammal often have hundreds of cell types, though our results in this paper can show coexistence with only few cell types.
The number of cell types in our dynamical differentiation model does not show clear increase with the number of chemical substances.
We suppose that the reason for these few types of cells
is due to our random choice of chemical reaction network, where the reaction paths are chosen equivalently.
In real biological system, the chemical reaction network is more organized, possibly in a hierarchical manner.

To choose such suitable network, evolutionary aspect of 
chemical reaction network should be taken into account for our model.
This problem also concerns with the emergence of gene expression system, 
and is under our current investigation.
\\\\
{\sl acknowledgements}
\\\\
The authors are grateful to  T. Yomo, T. Ikegami, S. Sasa, and T. Yamamoto
for stimulating discussions.
The work is partially supported by
Grant-in-Aids for Scientific
Research from the Ministry of Education, Science, and Culture
of Japan.

\pagebreak

Figure caption
\\\\
Fig.1: Overlaid time series of $x^{(m)}(t)$ of a single cell in medium, obtained form a network with 16 chemicals and three connections in each chemical.
Only the time series of
5 chemicals are plotted out of 16 internal chemicals.
Each line with the number {\it m}=2,9,10,11,12 gives the time series of the concentrations of the corresponding chemical $x^{(m)}(t)$.
This oscillatory behavior is a limit cycle, whose period {\it T} is longer than 
the plotted range of the figure (${\it T} \cong 16000$ time steps). 
The parameters are set as $e_1=1$, $D=0.01$, $f=0.01$,$\overline{X^{(\ell)}}=0.1$ for all $\ell$, and $V=100$. 
Chemicals $x^{(\ell)}(t)$ for $m \le 3$ are penetrable(i.e., $\sigma_{\ell}=1$), and others are not. 
The reaction network $Con(i,j,\ell)$ is randomly chosen initially, and is fixed 
throughout the simulation results of the present paper.
\\\\
Fig2:  Time series of $x^{(m)}(t)$, overlaid for the 5 chemicals 
(as given in Fig.1) in a cell.
(a)-(e) represent the course of differentiation to type-1, 2, 3, 4, and 5 cells respectively.
The differentiation to type-3, 4, and 5 cells always occurs from 
type-1 cells.
\\\\
Fig.3:  Orbits of internal chemical dynamics in the phase space.
(a) and (b) show the orbits of chemical concentrations for
a switching process from type-0 to type-1,2 cells,
respectively, plotted  in the projected space $(x^{(2)}(t),x^{(13)}(t))$.
Fig.3(c) gives a plot of $(x^{(1)}(t),x^{(8)}(t))$, which shows a switch 
from type-1 to type-3 cells.
(Note the difference of scales.)
Each cell type is clearly distinct in the phase space. 
\\\\
Fig.4:  Orbits of internal dynamics for each cell type.
The dynamics of each cell type is plotted in the same projected space 
as in Fig.3(a),(b), i.e., $(x^{(2)}(t),x^{(13)}(t))$.
Color corresponds to each cell type.
\\\\
Fig.5:  Cell lineage diagram.
Differentiation of cells (whose indices are given by the horizontal axis)
is plotted with time as the
vertical axis.
In this diagram, each bifurcation of lines through the horizontal segments
corresponds to the division of the cell, while the color indicates the 
cell type ( red for type 0, green for type 1, blue for type 2 respectively.).
We have plotted up to the stage of the differentiation to three types, while
the bifurcation to 3,4, and 5 from 1 occurs at a later course.
\\\\
Fig.6: Automaton-like representation of the rule of differentiation.
The path returning to the node itself represents the reproduction of its type, 
while the paths to other nodes represent the potentiality to differentiation 
to the corresponding cell types.
The dotted line from type-2 to type-0 gives an exceptional case:
Indeed the differentiation from ``2" to ``0" never occurs when
several types of cells such as  ``0",``1" and ``2" coexist. 
It occurs exceptionally only if
``5" cells dominate the system, when all cells are finally
differentiated to type-"5". In this case the type-``2" cells 
de-differentiate to ``0" ( and finally to  ``5").
\\\\
Fig.7: Variation of dynamics of type-2 cells with the change of the
rate of type-0 cell.
The distance $D_{2,2_0}$ of eq. (6) is measured between two type-2 cells
from the conditions $(n_0,n_2,n_3)=(23,50,27)$ and $(n_0,n_2,n_3)=(n_0,50,50-n_0)$,as one of type-3 cells is successively switched to type-0 externally by
$2\times10^4$ step.
Besides the distance, the number of type-0 cells is plotted against time.
\\\\
Fig.8:  Rate of the differentiation from type-0 to other cell types.
The total cell number is fixed  to 100
(without division process), while we take the initial cell distribution of
three types as $(n_0,n_1,n_2)= (n_0,30,70-n_0)$.
Starting the simulation with
this initial condition, the final number of each cell type,
as well as the number of differentiations from type-0 to others,
is plotted, as a function of the initial number of type-0 cells $(n_0)$.
Within this range of cell type distribution,
none of type-3,4,5 cells appear(see section 6).
\\\\
Fig.9:  Histogram about the number of type-2 cells.
Starting from a single cell with randomly chosen chemical concentrations,
the simulation is carried out until the total cell number reaches
300, when the number $n_2$ of type-2 cells is measured.    
Repeating the runs 347 times, we have counted the number of such initial conditions that $n_2$ falls onto a given bin (with the size 5).
The histogram
of $n_2$ is obtained from the count.
There are four peaks at $n_2$=0,100,150,220, each of which corresponds to 
a stable distribution of the cell colony.
\\\\
Fig.10:  Flow chart of the change of $(n_0,n_1,n_2)$.
We have carried out the simulations starting from the initial condition
at each $(n_0,100-n_0-n_2,n_2)$ by fixing the total cell number to 100 
(by removing the cell division process).  Change of the number of cell types is
measured from simulations, from which the direction of changes of $(n_0,n_2)$ is
shown as an arrow in the $(n_0,n_2)$ space. As is seen,
there are 5 fixed points, each of which corresponds to the stable population 
distribution of cell types.
\\\\

Table Caption
\\\\
Table I:  The average distance in phase space between each cell type:
$D_{i,j}$ in eq.(6) is estimated by taking the average  over $5\times10^4$ time steps.
Each cell type is sampled from a course of the evolution starting from
one cell.
\\\\
Table II:  Minimum distance in phase space between each cell type.
$D^{min}_{i,j}$ in eq.(7) is estimated from $5\times10^4$ time steps.
Each cell type is sampled from a course of  the evolution starting from
one cell.
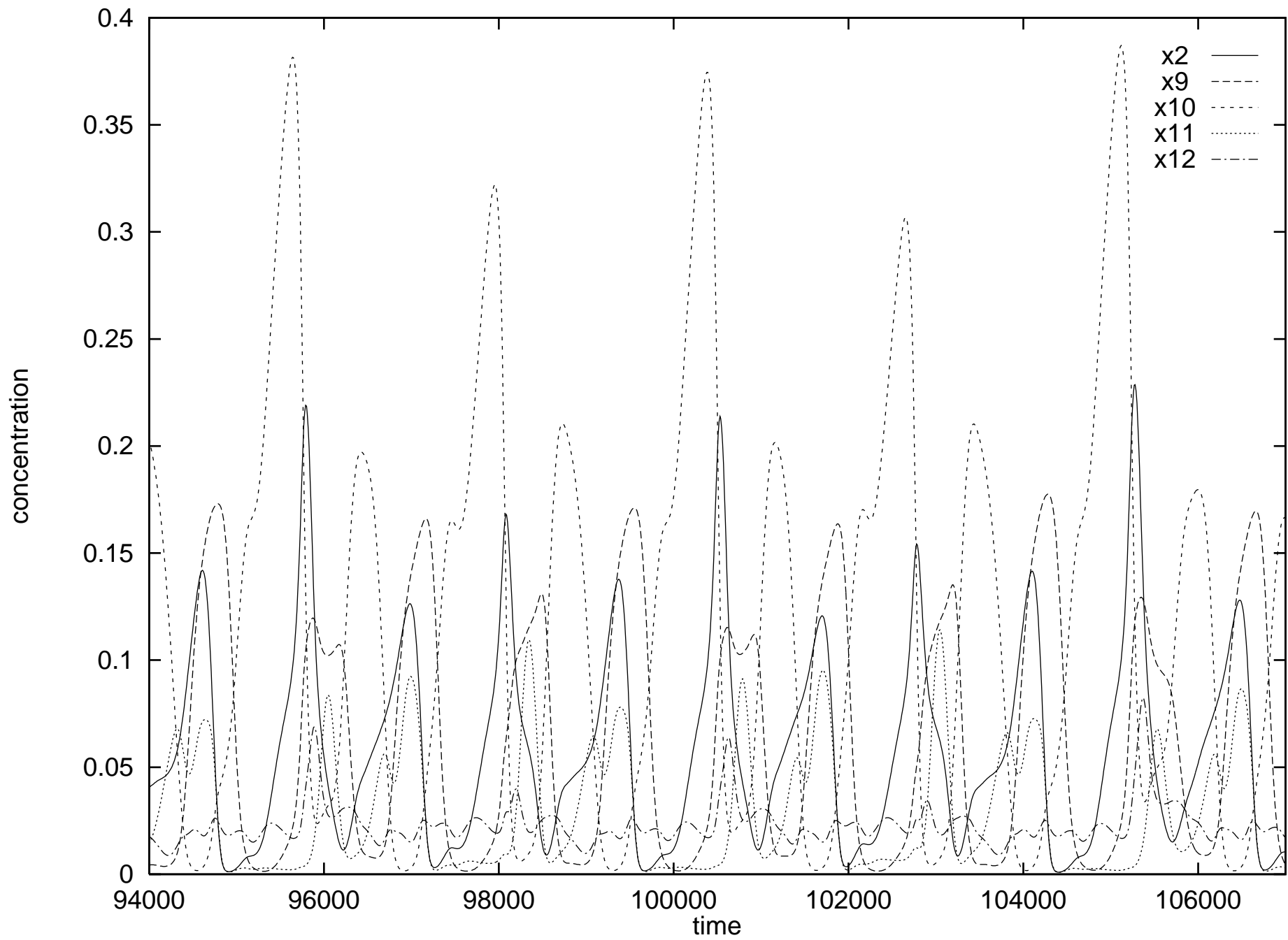
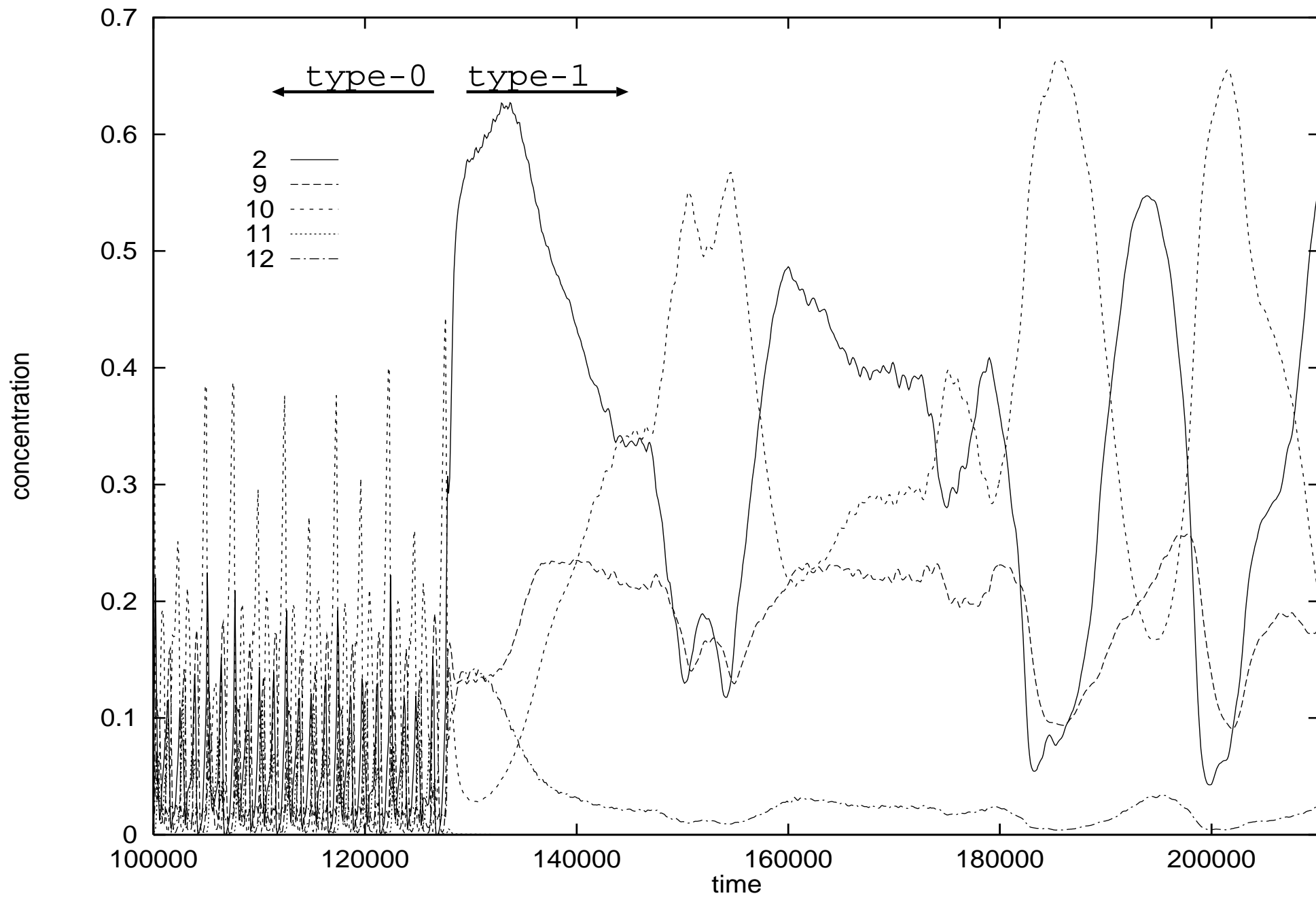
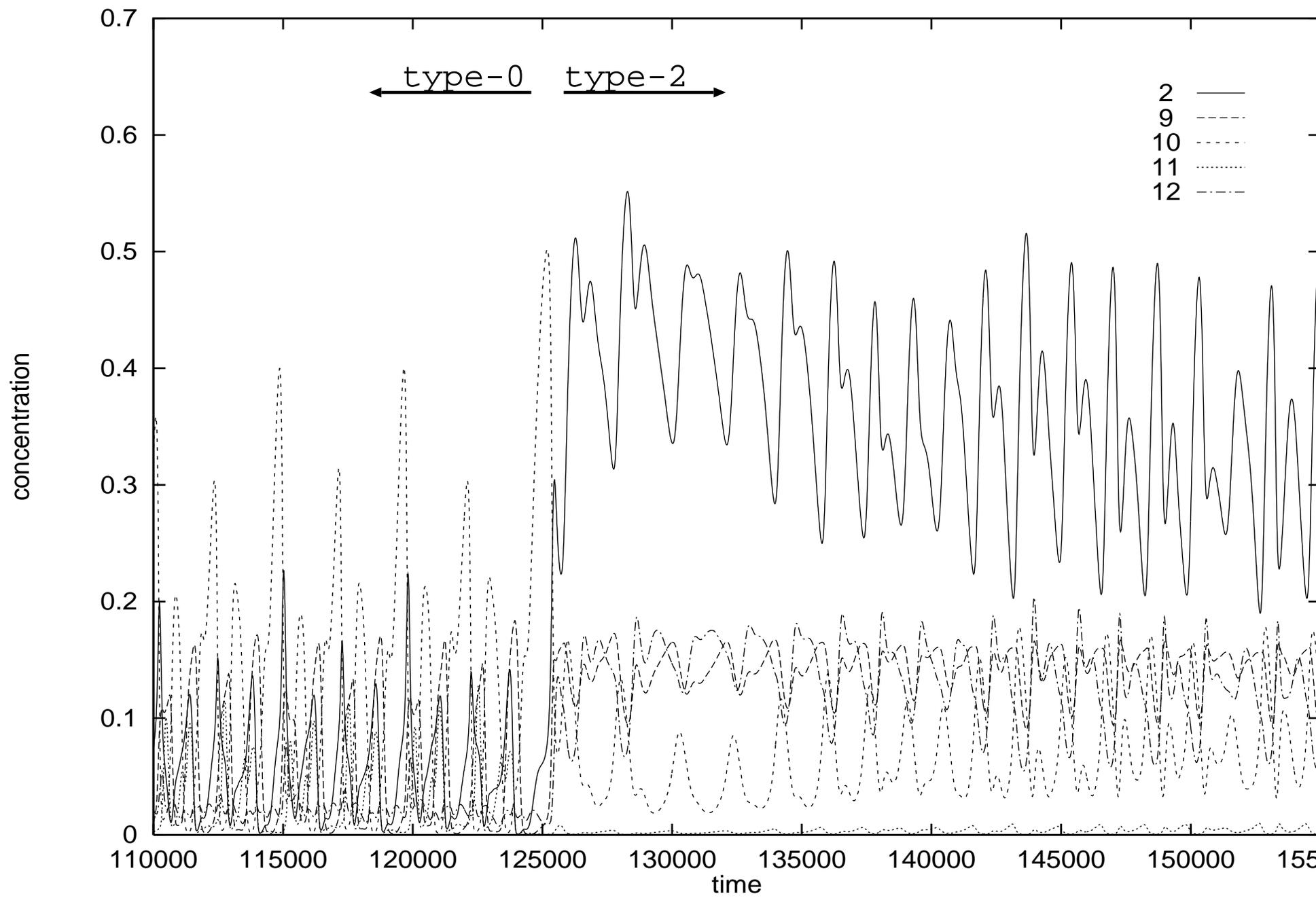
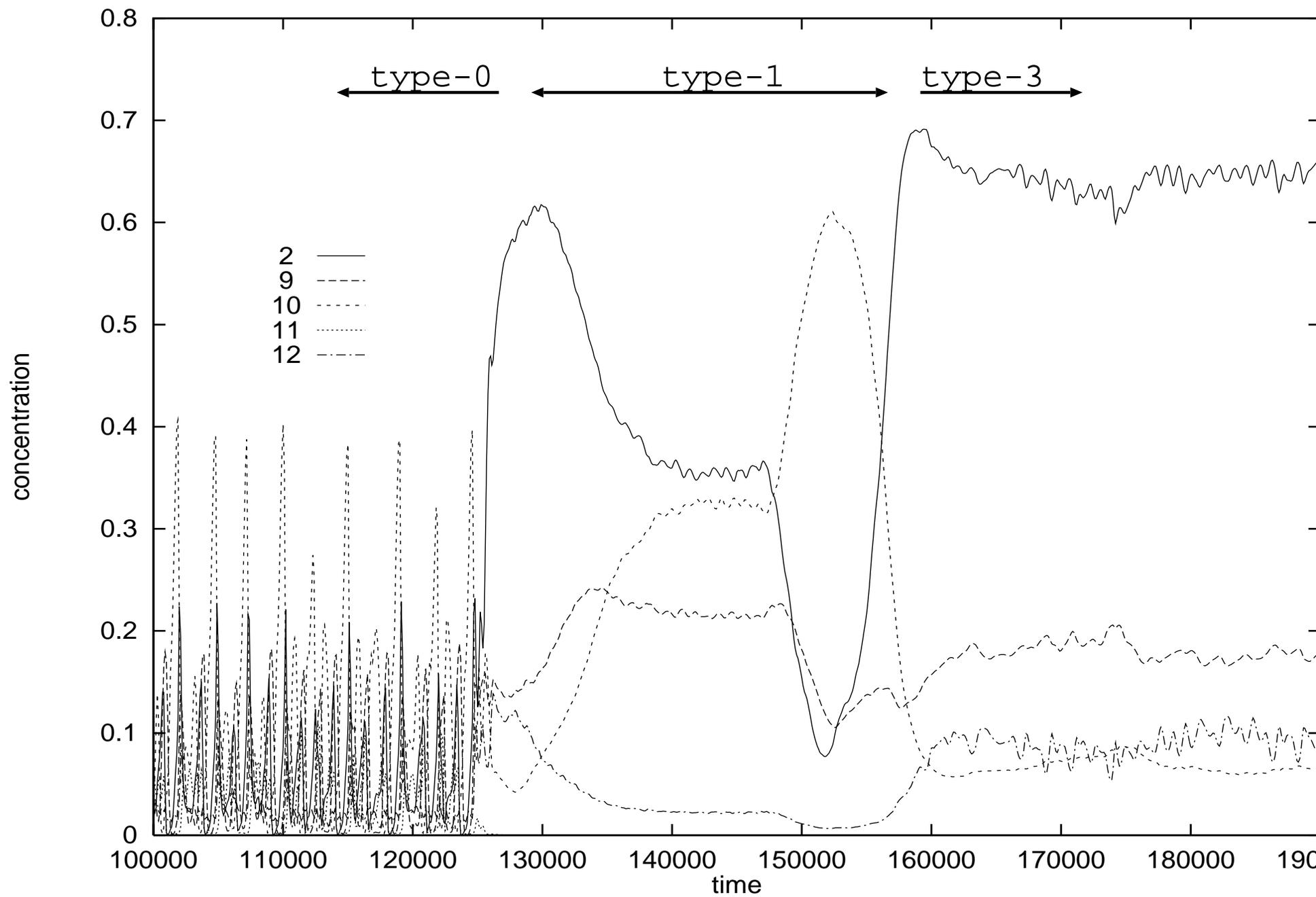
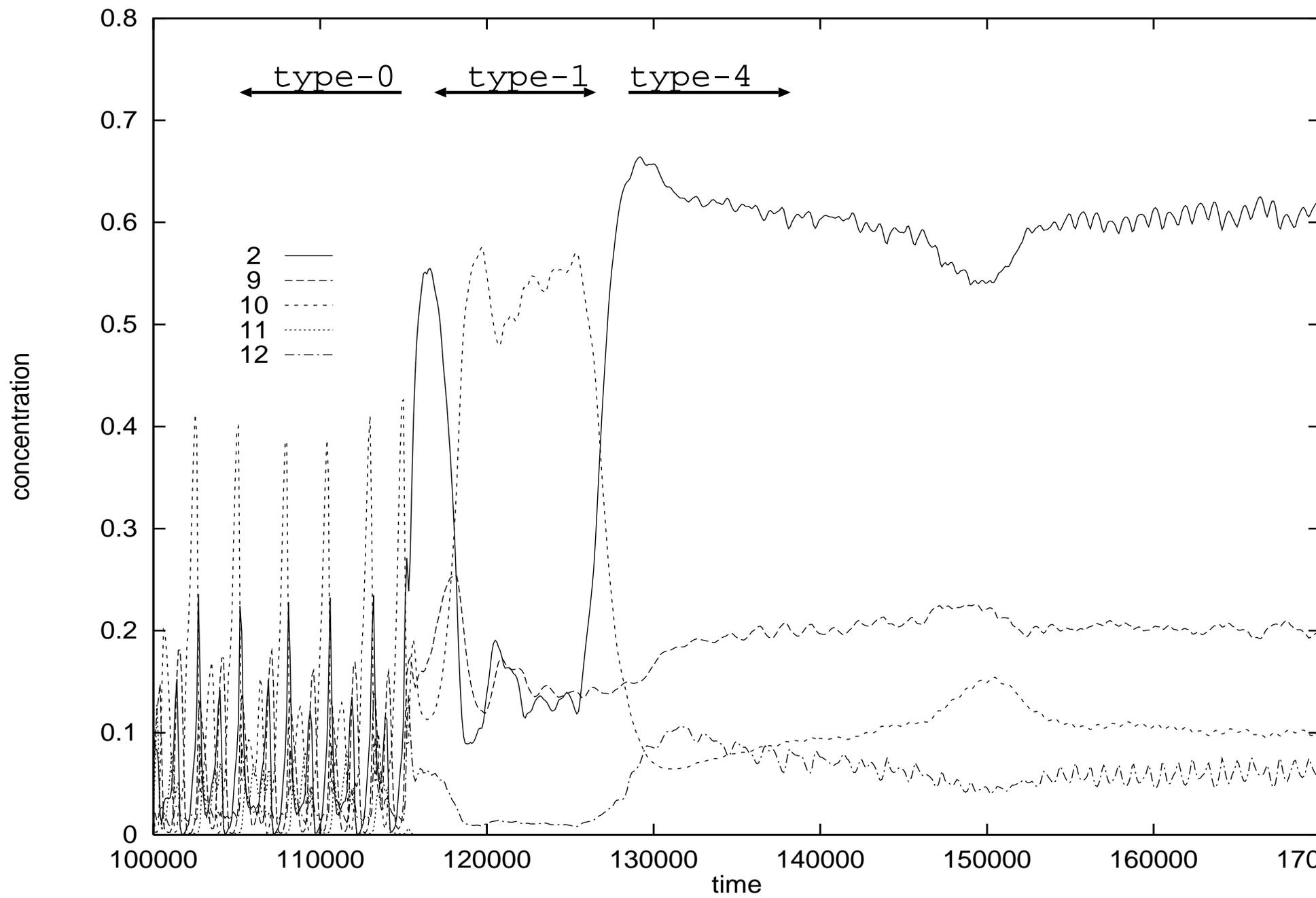
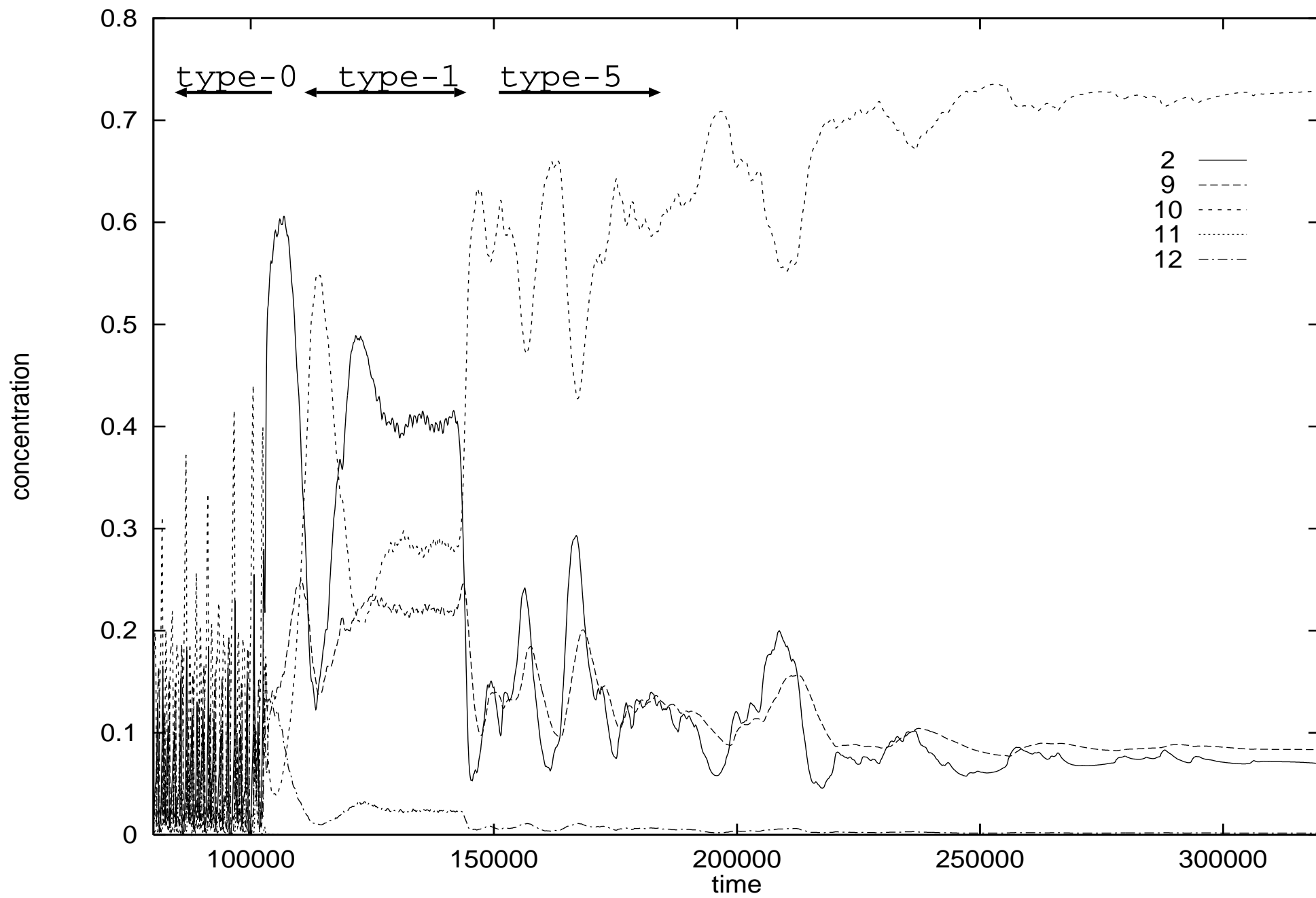
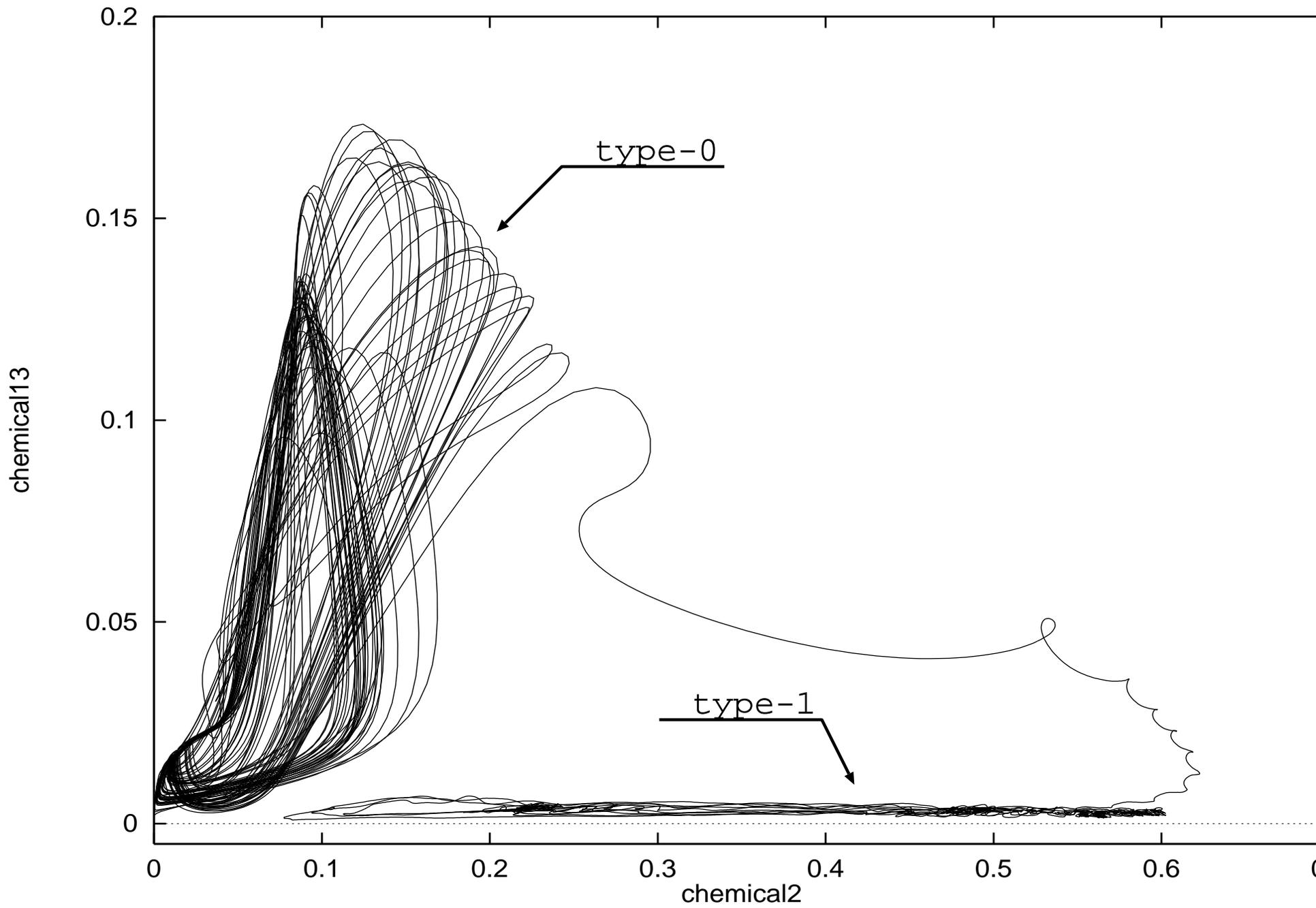
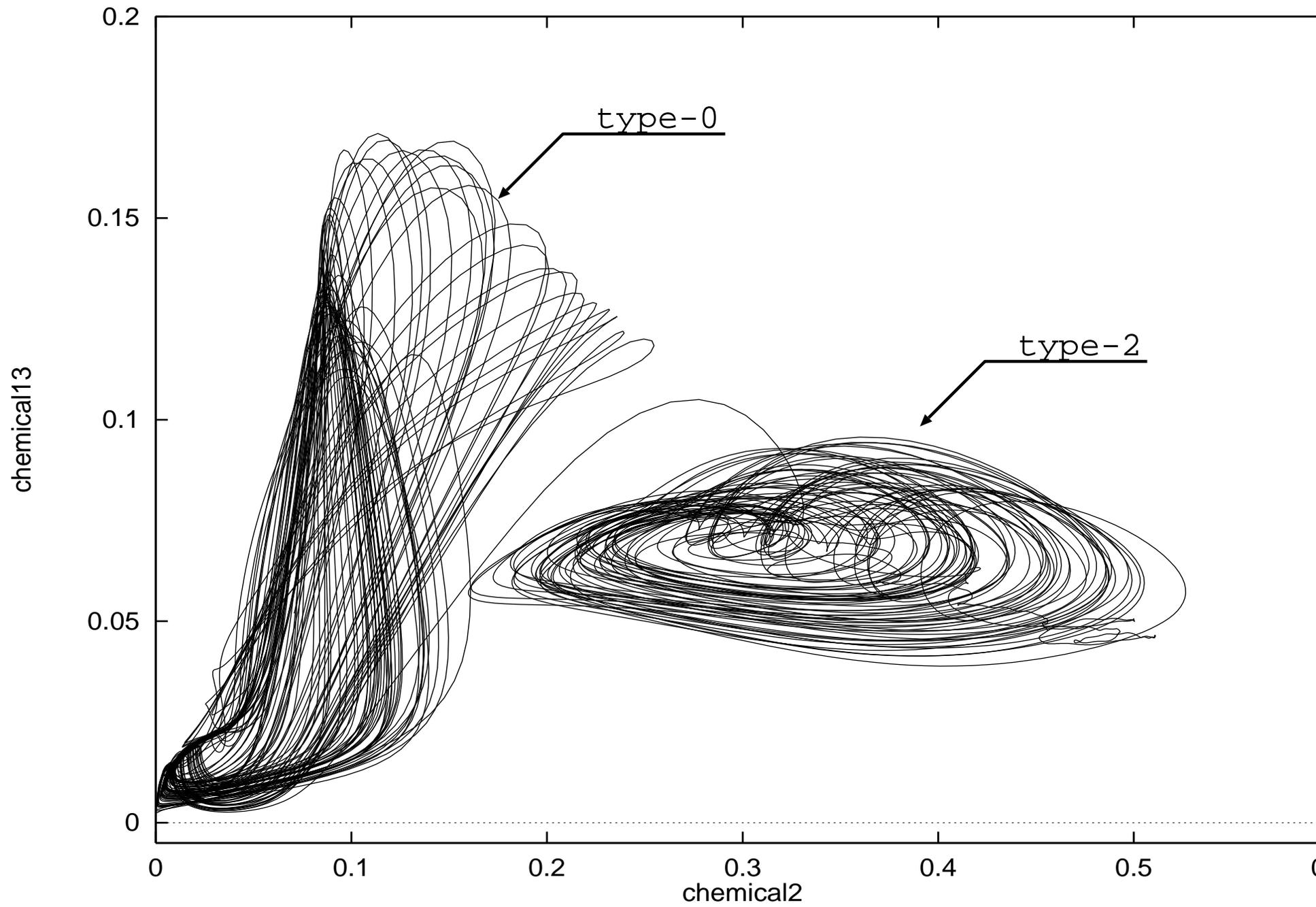
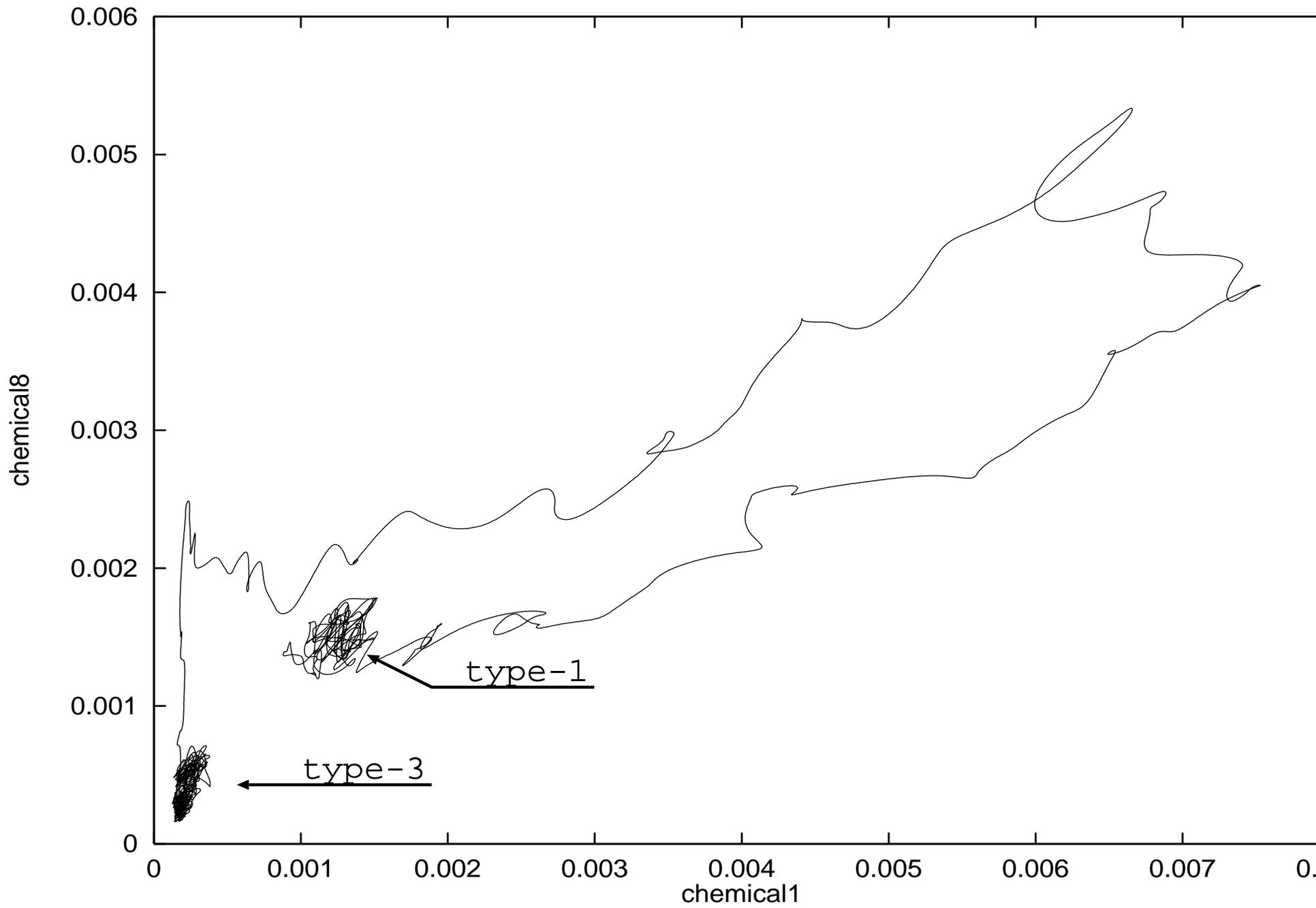
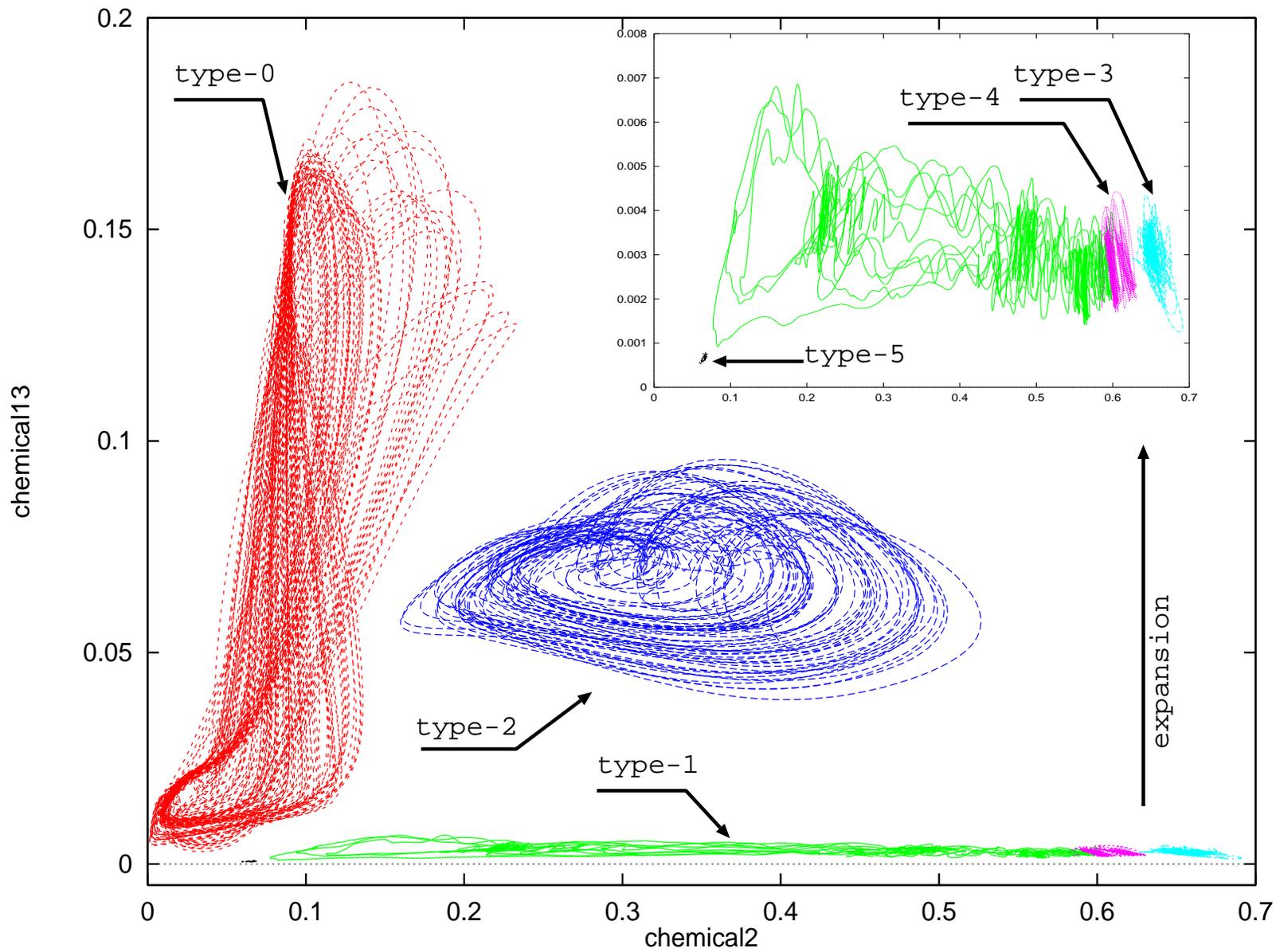
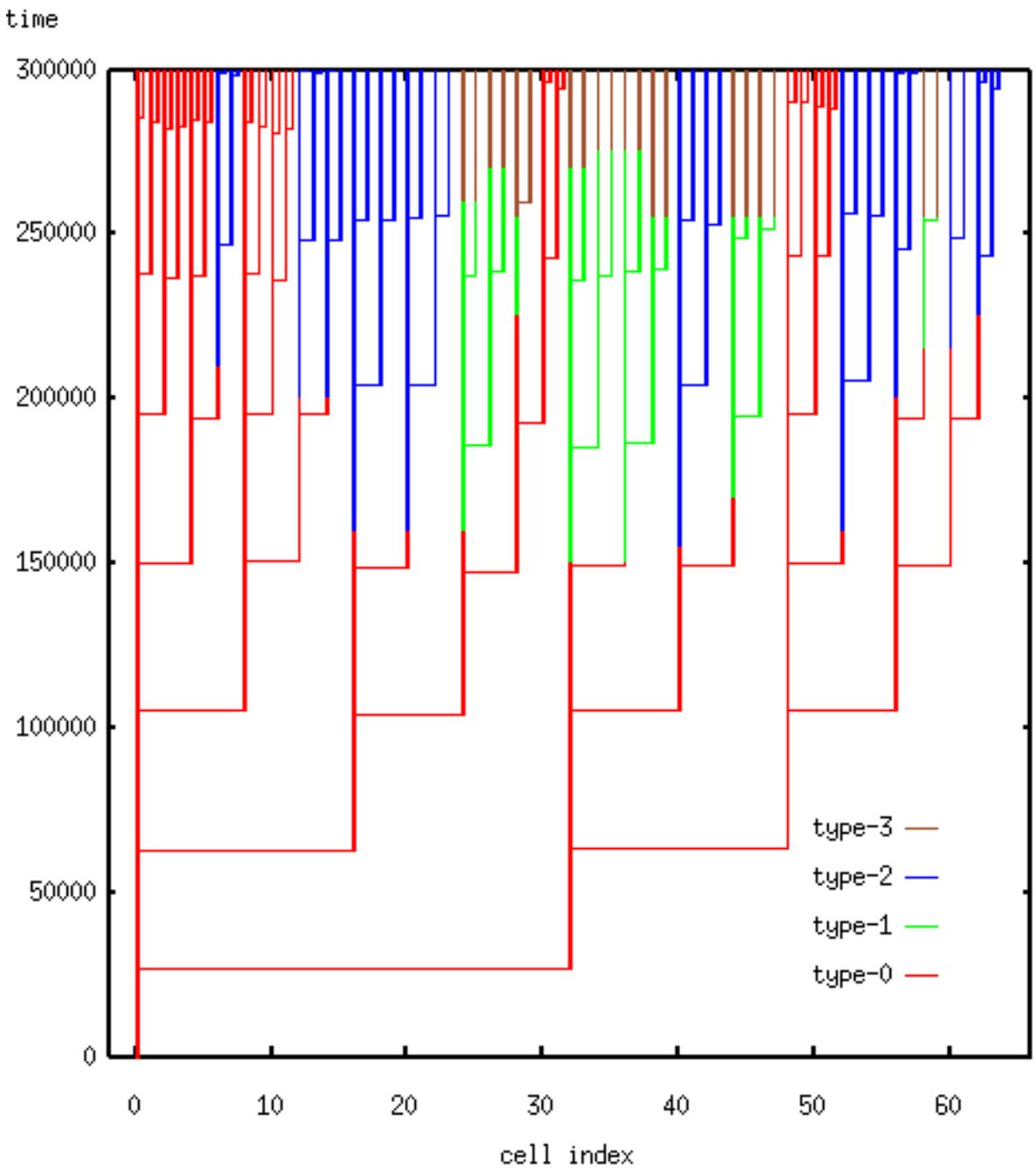
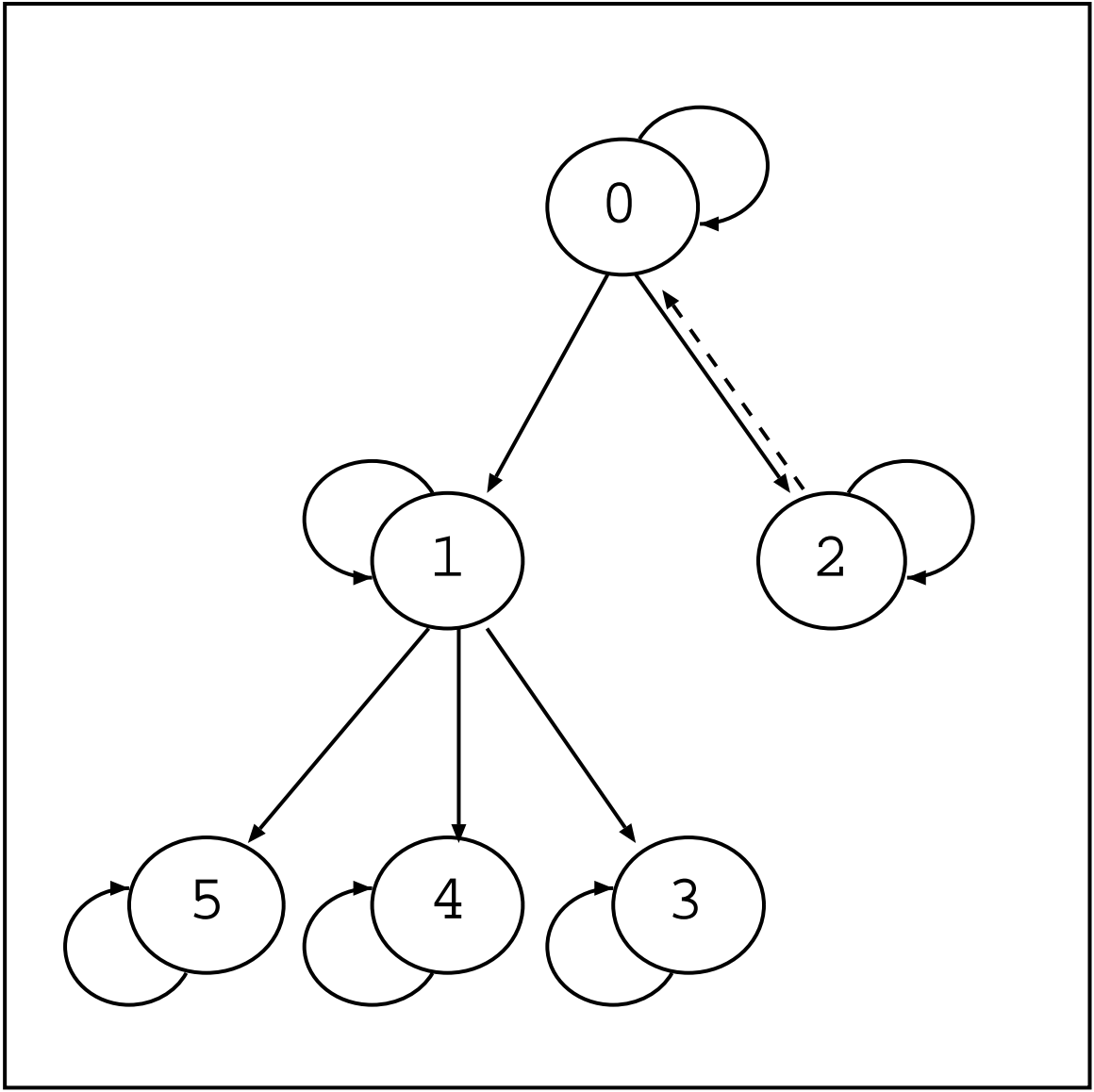
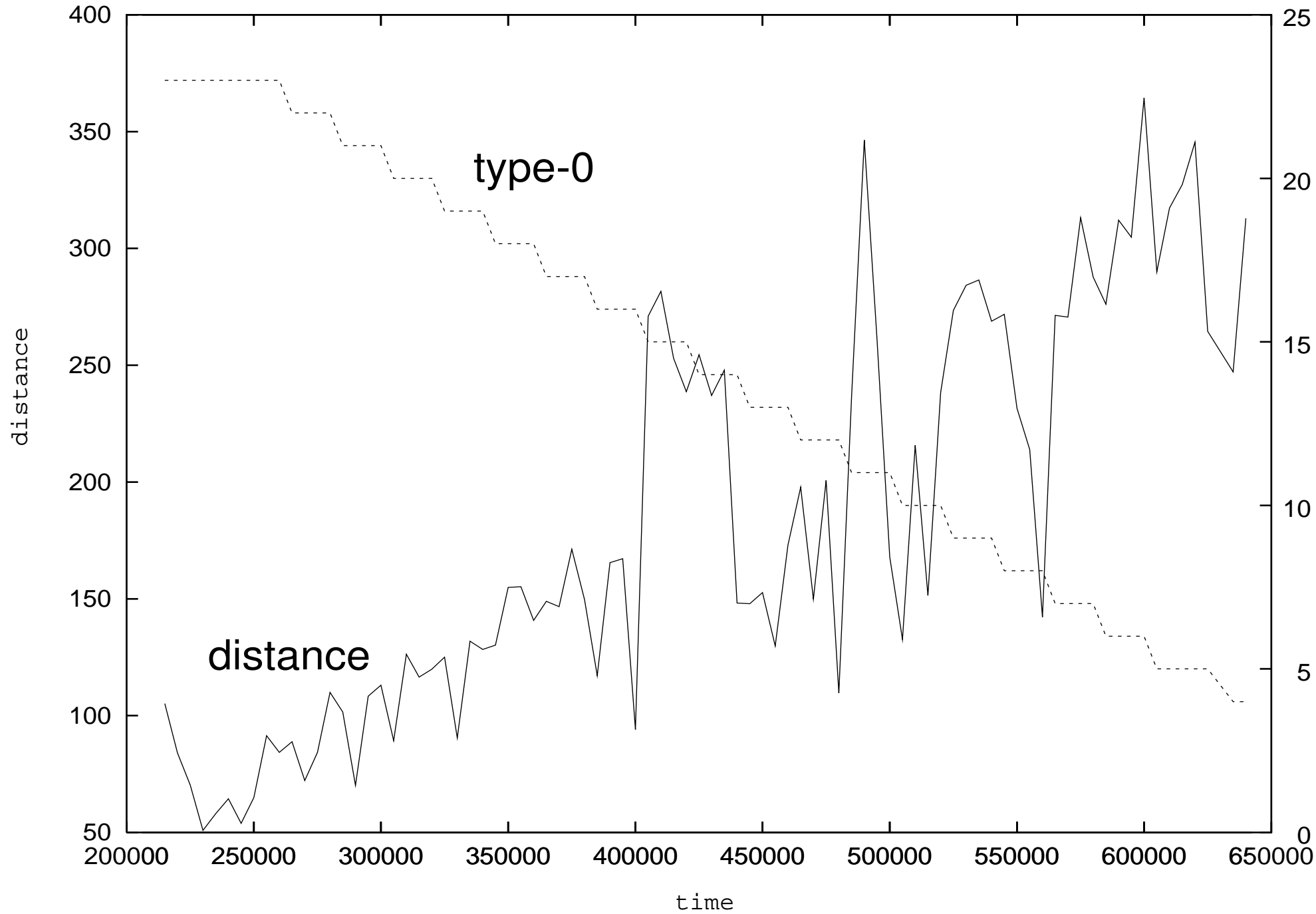
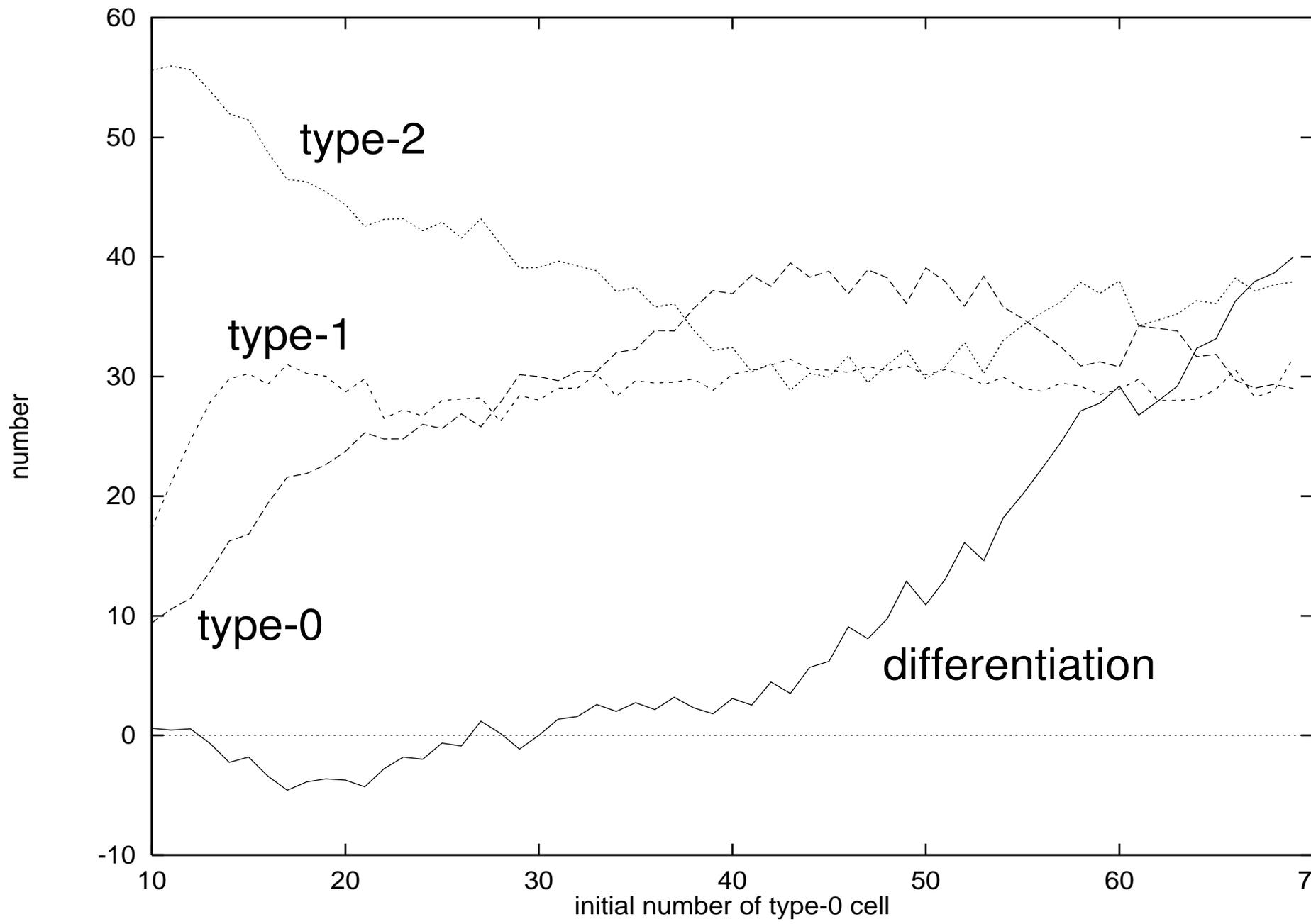
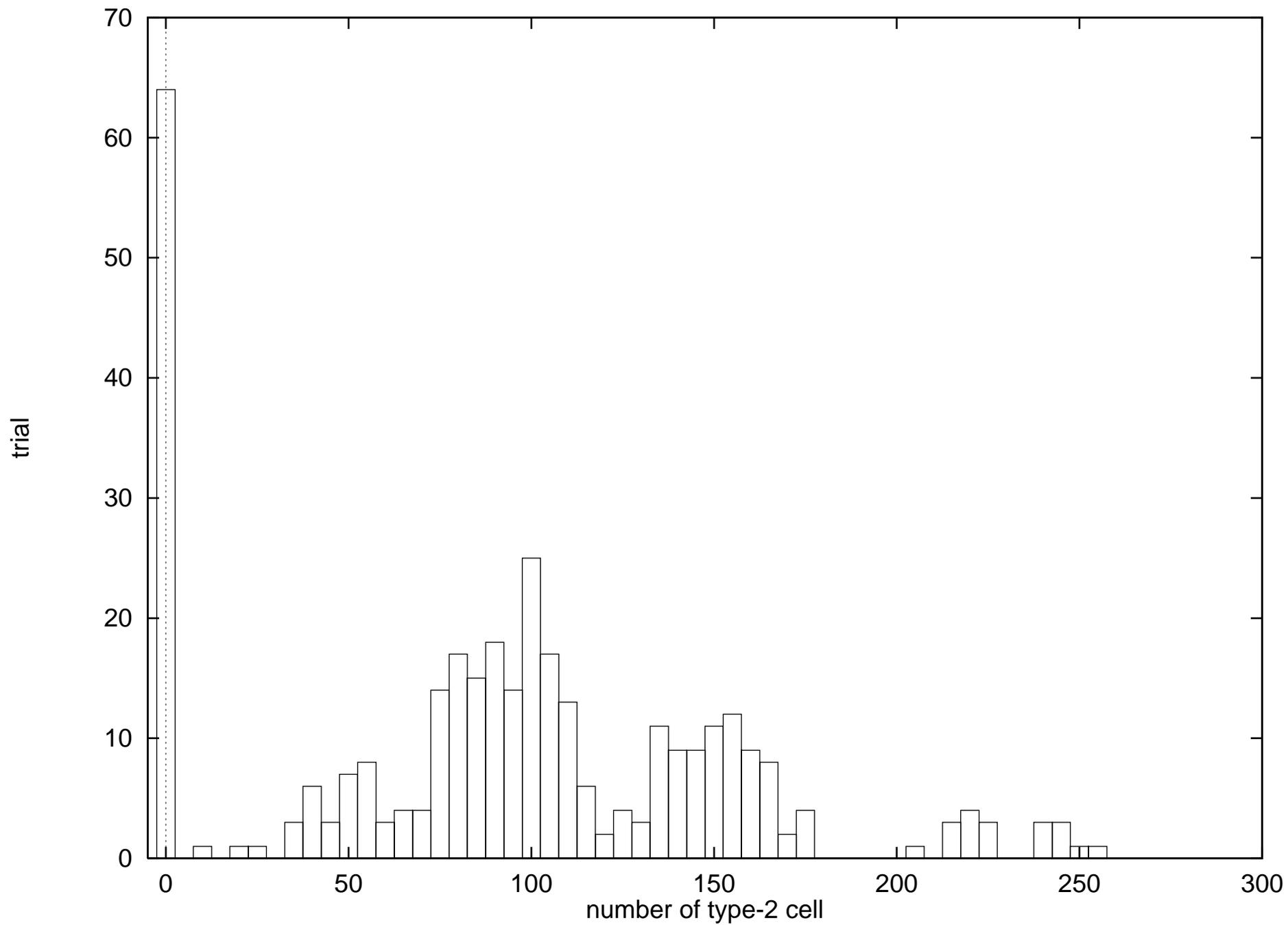
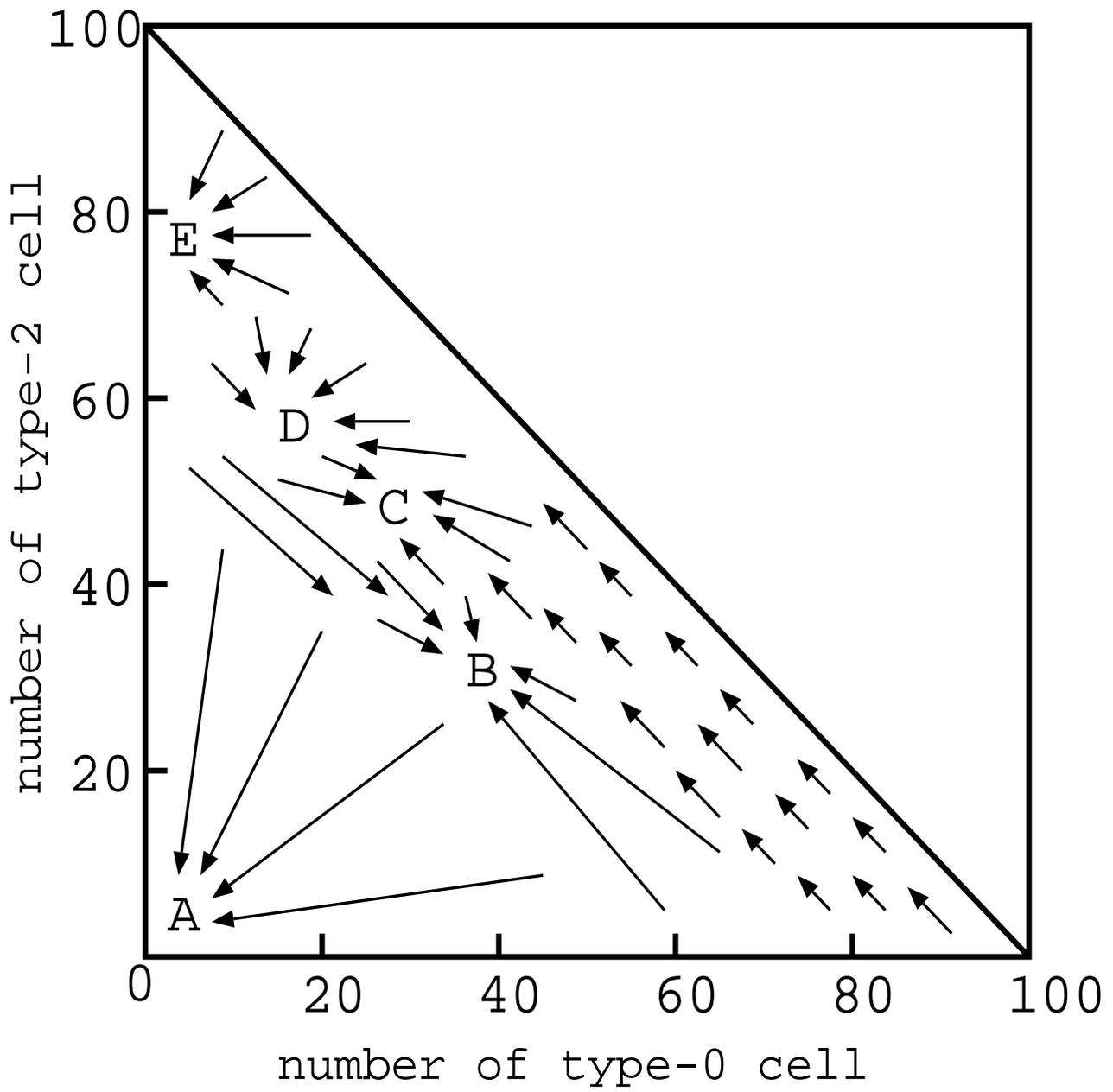
\end{document}